\documentclass[journal,comsoc]{IEEEtran}
\usepackage{array}
\usepackage{colortbl}
\usepackage[english]{babel}
\usepackage[T1]{fontenc}
\usepackage{changepage}
\ifCLASSINFOpdf

\else

\fi

\usepackage{amsmath}
\usepackage{supertabular}
\usepackage{comment}
\usepackage{cite}
\usepackage{color}
\usepackage{graphicx}
\usepackage{epsfig,graphics,subfigure,psfrag,amsmath,amssymb}
\usepackage{stfloats}
\usepackage{amsfonts}
\usepackage{slashbox}
\usepackage{multirow}
\usepackage{makecell}
\usepackage{amssymb}
\usepackage{algorithmic}
\usepackage{url}
\usepackage{hyperref}
\usepackage{breakurl}
\usepackage{bm}
\usepackage{mathrsfs}
\usepackage{amssymb}
\usepackage{makecell}
\usepackage{array}
\usepackage{textcomp}

\begin{document}


\title{IEEE 802.11be \--- Wi-Fi 7: New Challenges and Opportunities}

\author{Cailian~Deng\footnotemark{*},
        Xuming~Fang,~\IEEEmembership{Senior~Member,~IEEE,}
        Xiao~Han\footnotemark{*},
        Xianbin~Wang,~\IEEEmembership{Fellow,~IEEE,}
        Li~Yan,~\IEEEmembership{Student Member,~IEEE,}
        Rong~He,
        Yan~Long,~\IEEEmembership{Member,~IEEE,}
        and Yuchen~Guo
\thanks{Manuscript received XXX XX, 2019; revised XXX XX, XX.}
\thanks{The work of C. Deng, X. Fang, L. Yan, and R. He was supported in part by NSFC and High-Speed Rail Joint Foundation under Grant U1834210, Sichuan Provincial Applied Basic Research Plan under Grant 2020YJ0218, and Huawei HIRP Flagship Project under Grant HF2017060002. The work of Y. Long was supported in part by NSFC under Grant 61601380. \emph{(Corresponding author: Xuming Fang.)}}
\thanks{C. Deng, X. Fang, L. Yan, R. He, and Y. Long are with the Key Laboratory of Information Coding and Transmission, Southwest Jiaotong University, Chengdu 611756, China (e-mail:dengcailian@my.swjtu.edu.cn; xmfang@swjtu.edu.cn; liyan@swjtu.edu.cn; rhe@swjtu.edu.cn; yanlong@swjtu.edu.cn).}
\thanks{X. Han and Y. Guo are with the WT Laboratory, Huawei, Shenzhen 518129, China (e-mail: tony.hanxiao@huawei.com; guoyuchen@huawei.com).}
\thanks{X. Wang is with the Department of Electrical and Computer Engineering, University of Western Ontario, London, ON N6A 5B9, Canada (e-mail: xianbin.wang@uwo.ca).}
        }

{}

\maketitle
\renewcommand{\thefootnote}{\fnsymbol{footnote}} %
\footnotetext[1]{Co-first author.} %
\begin{abstract}
With the emergence of 4k/8k video, the throughput requirement of video delivery will keep grow to tens of Gbps. Other new high-throughput and low-latency video applications including augmented reality (AR), virtual reality (VR), and online gaming, are also proliferating. Due to the related stringent requirements, supporting these applications over wireless local area network (WLAN) is far beyond the capabilities of the new WLAN standard -- IEEE 802.11ax. To meet these emerging demands, the IEEE 802.11 will release a new amendment standard IEEE 802.11be -- Extremely High Throughput (EHT), also known as Wireless-Fidelity (Wi-Fi) 7. This article provides the comprehensive survey on the key medium access control (MAC) layer techniques and physical layer (PHY) techniques being discussed in the EHT task group, including the channelization and tone plan, multiple resource units (multi-RU) support, 4096 quadrature amplitude modulation (4096-QAM), preamble designs, multiple link operations (e.g., multi-link aggregation and channel access), multiple input multiple output (MIMO) enhancement, multiple access point (multi-AP) coordination (e.g., multi-AP joint transmission), enhanced link adaptation and retransmission protocols (e.g., hybrid automatic repeat request (HARQ)). This survey covers both the critical technologies being discussed in EHT standard and the related latest progresses from worldwide research. Besides, the potential developments beyond EHT are discussed to provide some possible future research directions for WLAN.
\end{abstract}


\IEEEpeerreviewmaketitle

\section{Introduction}

\IEEEPARstart{S}{ince} its adoption in 1990s, WLAN continues its growth in market share over the years and is becoming more and more important for providing wireless data services by using Wi-Fi technology. Home, enterprise and hotspots are increasingly dependent on Wi-Fi technology as their main access network. According to a recent study from the Wi-Fi Alliance \cite{ref1}, more than 9 billion Wi-Fi devices, including personal computers, smartphones, televisions, tablets, sensors, and so on, are currently in use worldwide. Video traffic is the dominant traffic type over WLAN, and its throughput requirement will keep increasing due to the emergence of 4k and 8k video whose uncompressed data rate is up to 20Gbps. Meanwhile, new ultra-high throughput and stringent low-latency applications are also proliferating, such as VR or AR, gaming (e.g., latency lower than 5ms for online gaming), telecommuting, online video conference, and cloud computing. While the recently released IEEE 802.11ax puts more focus on the network performance and user experience of the high-dense deployment scenarios, meeting the above high-throughput and low-latency requirements is well beyond the capabilities of IEEE 802.11ax. To meet these new needs, IEEE 802.11 standard organization is going to release a new amendment standard IEEE 802.11be EHT beyond IEEE 802.11ax, namely Wi-Fi 7. To be consistent with the IEEE 802.11be proposals, we mainly use the terminology ``EHT'' for IEEE 802.11be in this article. IEEE 802.11 working group has established a task group in May 2019 and specified the scope of the new generation WLAN, i.e., enabling new PHY and MAC modes to support a maximum throughput of at least 30 Gbps, and using carrier frequency operation between 1 and 7.250 GHz while ensuring backward compatibility and coexistence with legacy IEEE 802.11 devices in 2.4, 5 and 6 GHz bands \cite{ref2}. To achieve these goals, challenges of EHT have been identified and some enhanced PHY and MAC technologies have been explored as potential solutions to overcome these challenges.
\subsection{PHY Enhancements for EHT}
\begin{figure*}[!htbp]
  \begin{center}
    \setlength{\abovecaptionskip}{0.cm}
    \scalebox{0.36}[0.36]{\includegraphics{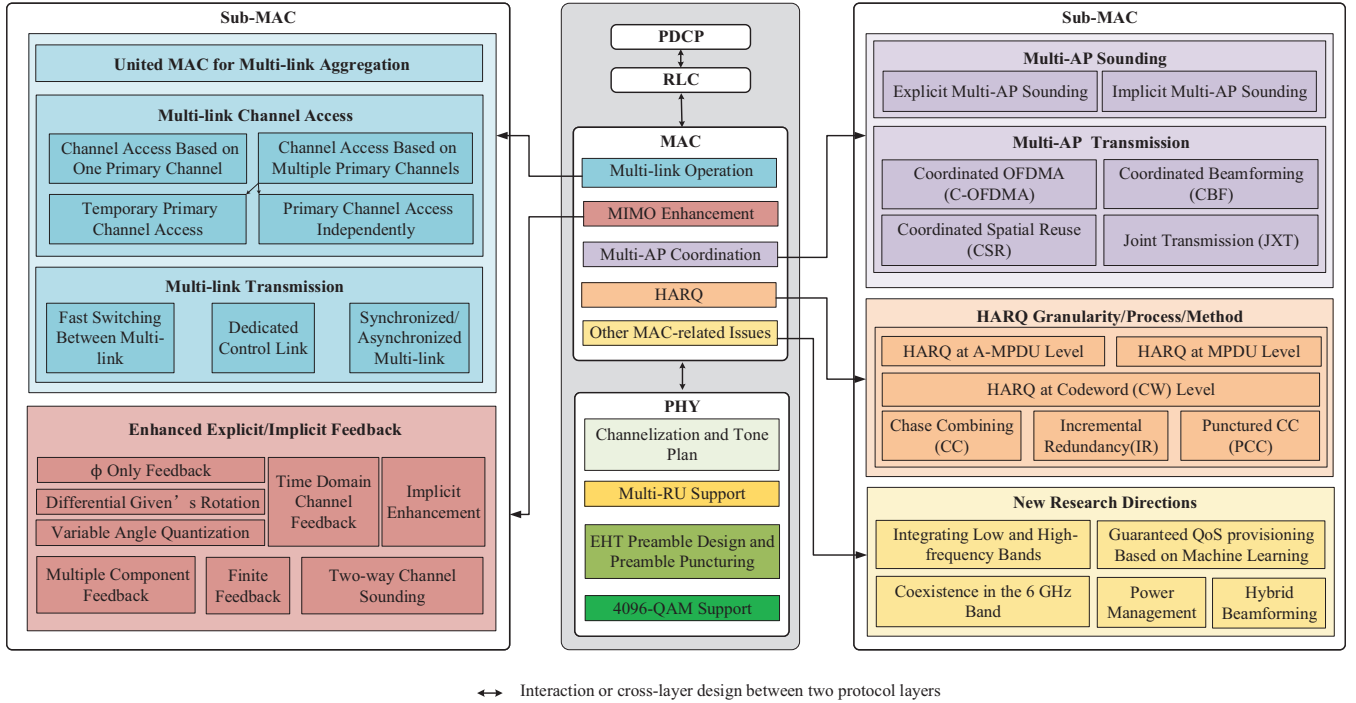}}
    \renewcommand{\figurename}{Fig.}
    \caption{Overview of the related technologies covered in this survey and their relationships.}
  \end{center}
\end{figure*}
\subsubsection{Providing expanded bandwidth of more than 160 MHz}
Due to the limited and crowded unlicensed spectra in 2.4 GHz and 5 GHz, the existing 802.11 WLANs (e.g., IEEE 802.11ax \cite{ref3}) will inevitably suffer from low Quality of Service (QoS) when running new emerging applications, such as VR/AR. To fulfill the promise of a maximum throughput of at least 30 Gbps, EHT is envisioned to add new bandwidth modes including contiguous 240 MHz, noncontiguous 160+80 MHz, contiguous 320 MHz and noncontiguous 160+160 MHz \cite{ref4}. Nevertheless, the channelization and tone plan for these new bandwidth modes are still under discussion, such as whether 240 MHz/160+80 MHz is formed by puncturing at 320/160+160 MHz, repeating the IEEE 802.11ax tone plan or defining a new tone plan for 160+160 MHz/320 MHz. Besides, EHT is supposed to design effective methods to improve the spectrum utilization of wideband and non-contiguous bandwidth.
\subsubsection{Supporting multi-RU assignment to a single user (SU)}
In IEEE 802.11ax, each user is only assigned to a specific RU for transmitting or receiving frames, which significantly limits the flexibility of the spectrum resource scheduling. To solve this problem and further enhance the spectral efficiency, the motion of allowing multi-RU assignment to a single user has been approved in the EHT task group \cite{ref4}. However, the related technical details are still pending in EHT, including multi-RU assignments, multi-RU combinations, coding and interleaving schemes for multi-RU, and signaling designs, and therefore more efforts in dealing with the relevant multi-RU issues are needed to realize the standardization of multi-RU in EHT.
\subsubsection{Introducing 4096-QAM for the peak data rate improvement}
The available highest-order modulation scheme of IEEE 802.11ax is 1024-QAM, where a modulated symbol carries 10 bits. To further improve the peak rate, 4096-QAM has been recommended for EHT to enable a modulated symbol to carry 12 bits. Therefore, given the same coding rate, EHT can gain a 20\% increase in data rate compared to 1024-QAM . Nevertheless, feasible configurations for 4096-QAM, such as coding strategies, the number of streams, error vector magnitude (EVM) requirements and multiple receiving antennas, still need to be explored and clarified for SU transmission and multi-user (MU) transmission modes.
\subsubsection{Providing efficient preamble formats and puncturing mechanisms}
Until EHT, different preamble formats have been introduced in each generation of WLAN standards, which can enable functions including synchronization, automatic gain control, time/frequency correction, channel estimation, auto-detection to differentiate the version of a physical protocol data unit (PPDU) and necessary signaling (e.g., resource allocation information), etc. According to the PAR \cite{ref5}, the EHT preamble design should ensure backward compatibility and coexistence with legacy PPDUs transmitted on 2.4 GHz, 5 GHz, and 6 GHz bands. Besides, bringing future compatibility to preambles starting with EHT was proposed in \cite{ref6}. Since in EHT many new features like multi-RU and MU-MIMO are still being considered, formats and details of the preamble for supporting different technologies and scenarios are still pending. By bonding channels in a non-continuous way, the new preamble puncturing in IEEE 802.11ax allows a Wi-Fi device to transmit the MU PPDU over the entire bandwidth (e.g., 80 MHz, 80+80 MHz or 160 MHz) except for the punctured preamble part. However, due to the absence of the SIG-B field and the puncturing-related signaling in the preamble in the SU PPDU, the SU PPDU is not allowed to employ preamble puncturing and must be transmitted over the entire available contiguous bandwidth. Thus, in EHT, there may be a need to improve the puncturing design for the MU PPDU and add the puncturing design for the SU PPDU.
\subsection{MAC Enhancement for EHT}

\subsubsection{Multi-link operation over dramatically increased bandwidth}
Due to the limited and crowded unlicensed spectra in 2.4 GHz and 5 GHz, the existing IEEE 802.11 WLANs (e.g., IEEE 802.11ax) suffer low QoS to serve new emerging application use cases, such as VR/AR. To fulfill the promise of a maximum throughput of at least 30 Gbps, EHT expands its bandwidth by multi-band aggregation across 2.4 GHz, 5 GHz and 6 GHz bands, gaining up to 320 MHz bandwidth. However, challenges such as channel frequency selectivity over a much broader and noncontiguous bandwidth, different types of multi-band operations and backward compatibility and coexistence with existing legacy STAs in 2.4 GHz, 5 GHz and 6 GHz bands will arise when multi-band aggregation is performed. In the legacy multi-band operations (e.g., fast session transfer (FST)\cite{ref7}), there is a limitation that MAC service data units (MSDUs) belonging to a single traffic identification (TID) can only use single band, resulting in significant MAC overheads for session transfer. Thus, to improve the transmission flexibility and minimize the MAC overhead, the existing MAC models may need a major improvement in EHT, that is, an STA can transmit frames of the same TID or different TIDs over multiple bands concurrently or non-concurrently. For such MAC enhancement, the terminology ``multi-kink'' used in EHT is preferred over the ``multi-band'' \cite{ref8}. However, to standardize multi-link support in EHT, discussions and efforts on multi-link architecture, operation, and functions still need to continue.
\subsubsection{Supporting increased spatial streams and MIMO enhancements}
To meet the growing traffic demands generated by the increasing number of Wi-Fi devices, APs have continued to increase the number of antennas and better spatial multiplexing capabilities over the recent years. Currently, in IEEE 802.11ax \cite{ref3}, an AP equipped with 8 antennas can simultaneously serve up to 8 users for uplink (UL)/downlink (DL) transmission, through MU-MIMO. Continuing the trend of upgrading AP's spatial multiplexing capability, EHT recommends the maximum spatial streams of 16 to gain higher network capacity. However, increasing the number of spatial streams comes with an attendant increase in the overhead of acquiring CSI (channel state information). With 16 spatial streams in EHT, reusing the same channel sounding method specified in the current IEEE 802.11ax will result in the enormous CSI feedback overhead. For this reason, EHT needs to improve existing explicit and implicit feedback schemes for overhead reduction or develop completely new CSI feedback schemes.
\subsubsection{Distributed operations among neighboring APs}
IEEE 802.11ax only supports transmission to/from a single AP and spatial reuse between APs and STAs without coordination among neighboring APs. As a result, its capability to utilize the flexibility of time, frequency and spatial resources is significantly limited. To improve this, EHT extends its capability to support sharing data and control information among APs via wired or wireless links, thus improving the spectrum efficiency, increasing the peak throughput and reducing the latency. This major feature differentiating EHT from IEEE 802.11ax is referred to as multi-AP coordination, which can be divided into coordinated spatial reuse (CSR), coordinated orthogonal frequency-division multiple access (C-OFDMA), coordinated beamforming (CBF) and joint transmission (JXT) according to different coordination complexity. The selection of multi-AP transmission modes is based on the scenario requirements. In a typical multi-AP network architecture (e.g., enterprise network) without a central node, an AP has to communicate with each neighboring AP for coordination, which will result in substantial signaling overhead and processing complexity. Therefore, an efficient coordination procedure (including the multi-AP sounding, the multi-AP selection and multi-AP transmission) with low overhead and processing complexity is needed to support all types of multi-AP coordination. Besides, precise phase/time synchronization and proper resource allocation functions are crucial to avoid mutual interference between neighboring APs, as imperfect synchronization may cause peak throughput degradation significantly.

\subsubsection{Enhanced link adaptation and retransmission mechanism}
Transmission reliability is also another major concern for EHT. Current IEEE 802.11 systems rely on the retransmission of MAC protocol data unit (MPDU) (s) to ensure transmission reliability in randomly varying and error-prone wireless channels. In the automatic repeat request (ARQ) protocol, the receiver simply discards the erroneous MPDU(s) before receiving its retransmitted MPDU(s). With the requirement on higher reliability and lower latency, HARQ is expected to implement in EHT, which enables soft combining or additional parity at the receiver to improve the likelihood of correct decoding. Unlike ARQ, in HARQ, the receiver will store incorrectly decoded packets and combine them with subsequent retransmissions before decoding. Nevertheless, several issues regarding implementation of HARQ in the 802.11-like system are raised, including retransmission granularity (e.g., Aggregate MPDU (A-MPDU), MPDU or codeword (CW)), HARQ process, HARQ method (e.g., chase combining (CC) or incremental redundancy (IR)), link adaptation methods for higher HARQ gains and so on. How and at which layer HARQ can be better supported as well as what changes would be necessary at the PHY and MAC layer are extremely challenging for EHT.
\begin{figure*}[!htbp]
  \begin{center}
    \setlength{\abovecaptionskip}{0.cm}
    \scalebox{0.7}[0.7]{\includegraphics{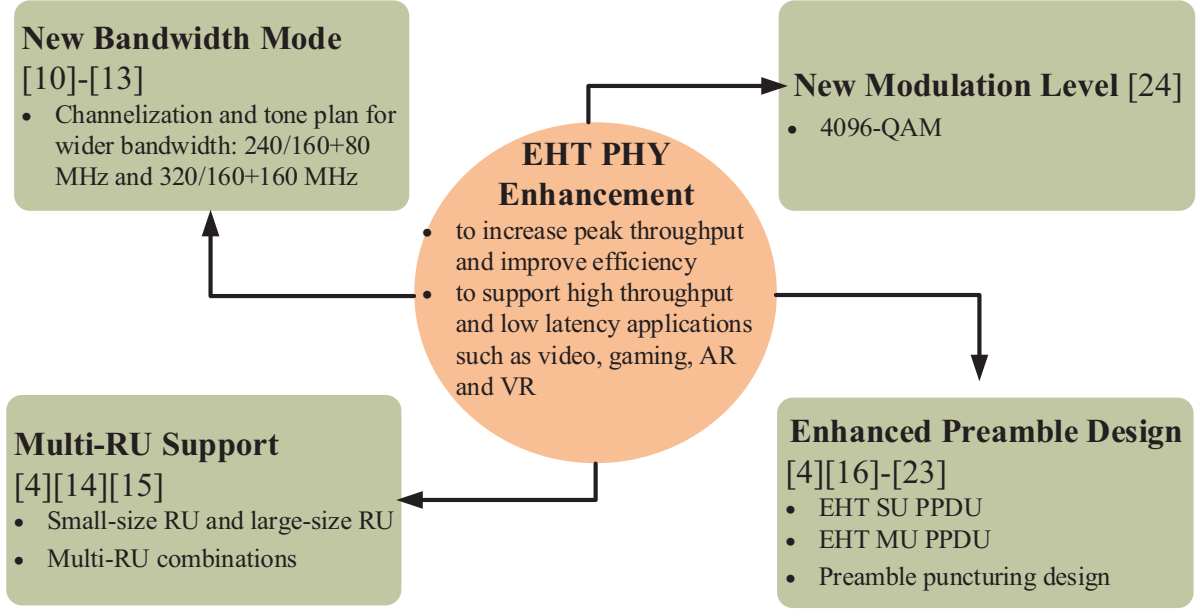}}
    \renewcommand{\figurename}{Fig.}
    \caption{Key PHY enhancements for EHT.}
    \label{fig_2}
  \end{center}
\end{figure*}
\par
The authors of this article are involved in the research and design of the EHT standards. To provide the comprehensive understanding of the EHT standardization activities to the readers, this article investigates the latest standardization progress of EHT during the task group phase as well as the new progress of the related academic studies. To the best of our knowledge, there is the first survey work on the development of the EHT technical specifications during the task group phase. A valuable article \cite{ref9} has surveyed the candidate technical features discussed in the EHT fora during the initial topic interest group and study group phases, provided system-level simulation results to evaluate the potential throughput gains, and discussed the coexistence issues with other technologies operating in the 6 GHz band. In this article, we focus on investigating what the corresponding techniques and solutions are proposed in the EHT task group to improve the network performance. To illustrate the structural relationship between each section of this article, Fig. 1 provides an overview of the survey. This article doesn't intend to cover the packet data convergence protocol (PDCP) layer and radio link control (RLC) layer, but focus its attention on new features in both PHY and MAC layers for EHT. As we can see in Fig. 1, the new important PHY related techniques for EHT are included, namely, channelization and tone plan, multi-RU support, 4096-QAM, preamble design and preamble puncturing. For the MAC related techniques, we mainly introduce multi-link operations, MIMO enhancement, multi-AP coordination, and HARQ. Besides, we put forward some new research directions and viewpoints beyond EHT for promoting the development of wireless communication networks.\par

\begin{figure*}[!htbp]
  \begin{center}
    \setlength{\abovecaptionskip}{-0.2cm}
    \scalebox{0.8}[0.8]{\includegraphics{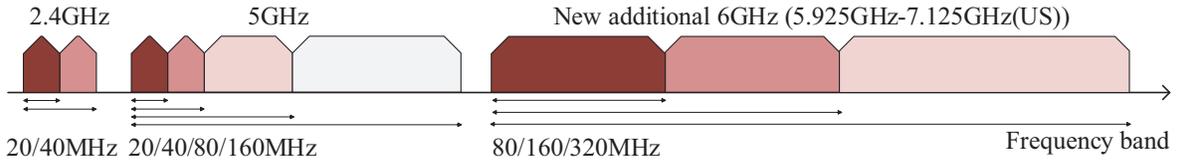}}
    \renewcommand{\figurename}{Fig.}
    {\color{black}\caption{Available transmission bandwidth over 2.4 GHz, 5 GHz and 6 GHz frequency bands. In the 2.4 GHz band, IEEE 802.11n devices are allowed to transmit in a 40 MHz channel by aggregating two adjacent 20 MHz channels into a single 40 MHz channel. To support high-speed wireless communication demands, IEEE 802.11ac/IEEE 802.11ax introduced the capability of extending the number of basic channels, thereby allowing mobile devices to transmit over an 80 MHz/160 MHz channel in the 5 GHz band. In the 6 GHz band, more than 1 GHz of additional unlicensed spectrum is available, allowing mobile devices to transmit in bandwidths up to 320 MHz.}}
  \end{center}
\end{figure*}
The remainder of this article is organized as follows: Section II presents the key PHY enhancement techniques. Section III-VII focus on MAC enhancements and cross-layer techniques between PHY and MAC layer. Specifically, Section III presents multi-link aggregation and operations. Section IV emphasizes on MIMO enhancement. Section V describes multi-AP coordination technologies. Section VI provides enhanced link adaptation and retransmission protocols. Section VII introduces potential perspectives beyond EHT. Finally, Section VIII concludes this article.\par
For presentation clarity, technical terms and acronyms used repeatedly in this article are listed alphabetically in Appendix A.
\section{PHY Enhancements for EHT}
To support high-throughput and low-latency video applications such as AR, VR and online gaming, EHT introduces several PHY enhancement technologies shown in Fig. 2, which enables the EHT to achieve an ultra-high peak rate of up to 30 Gbps. And PHY enhancements and related works are summarised in Table I.
\par(1) Wider bandwidth modes including 320 MHz, 160+160 MHz, 240 MHz and 160+80 MHz, have been identified as one of the candidate features in capacity augments for EHT.
\par(2) Multi-RU assigned to a single user is supported to enhance the spectral efficiency in EHT.
\par(3) EHT recommends new higher-order modulation strategies, namely 4096-QAM, to further increase the peak rate compared to 1024-QAM adopted in IEEE 802.11ax.
\par(4) Two EHT preamble formats are being considered for EHT SU PPDU and EHT MU PPDU, respectively. To improve the spectral efficiency, EHT PHY is also supposed to adopt a new preamble puncturing mechanism for an EHT PPDU transmitted to one or more STAs.\par
In this section, we will discuss the aforementioned PHY enhancements in more detail.
\begin{table*}[!htbp]
\renewcommand\arraystretch{1.2}
\centering
{\color{black}\caption{Summary of PHY Enhancements and Related Works.}}
\begin{tabular}{|c|l|}
\hline
{\color{black} \textbf{PHY Enhancement}}                       & \multicolumn{1}{c|}{{\color{black} \textbf{Contributions}}}                                                                                                                                                                                                                                                                                                       \\ \hline
{\color{black} }                                               & {\color{black}\-- Way forward on IEEE 802.11be specification development for 6 GHz band support {[}10{]}}                                                                                                                                                                                                                                                 \\ \cline{2-2}
{\color{black} }                                               & {\color{black}\-- Discussion on multi-band operation, flexible channel aggregation {[}11{]} {[}12{]}}                                                                                                                                                                                                                                                               \\ \cline{2-2}
\multirow{-3}{*}{{\color{black}\textbf{New Bandwidth Mode}}}  & {\color{black} \begin{tabular}[c]{@{}l@{}}\-- Channelization of 320 MHz, EHT PPDU bandwidth modes and EHT tone plan designs for 320 MHz {[}13{]}\end{tabular}}                                                                                                                                                                        \\ \hline
{\color{black} }                                               & {\color{black}\-- Summary of proposal development of resource unit, such as single RU, multi-RU, coding, etc. {[}4{]}}                                                                                                                                                                                                                                              \\ \cline{2-2}
{\color{black} }                                               & {\color{black} \begin{tabular}[c]{@{}l@{}}\-- Maximum number of RUs assigned to a single STA and restrictions on the combination and locations \\~{} of RUs {[}14{]}\end{tabular}}                                                                                                                                                                        \\ \cline{2-2}
\multirow{-3}{*}{{\color{black} \textbf{Multi-RU Support}}}    & {\color{black} \begin{tabular}[c]{@{}l@{}}\-- Discussion on several aspects regarding multiple RUs for one user transmission: PPDU format, transmission\\ ~{} in Data field and signaling {[}15{]}\end{tabular}}                                                                                                                                                        \\ \hline
{\color{black} }                                               & {\color{black} \begin{tabular}[c]{@{}l@{}}\-- Three phase rotation design options for 320 MHz: repeat IEEE 802.11ax/new phase rotation, repeat\\ ~{} IEEE 802.11ax/new phase rotation and apply additional phase rotation, and find optimal phase\\  ~{} rotation {[}16{]}\end{tabular}} \\ \cline{2-2}
{\color{black} }                                               & {\color{black}\-- Preamble structure designs {[}17{]} {[}18{]}}                                                                                                                                                                                                                                                                                                      \\ \cline{2-2}
{\color{black} }                                               & {\color{black} \begin{tabular}[c]{@{}l@{}}\-- EHT P matrices design for EHT-LTF and the new considered dimension of P matrix {[}19{]}\end{tabular}}                                                                                                                                                                                                  \\ \cline{2-2}
{\color{black} }                                               & {\color{black} \begin{tabular}[c]{@{}l@{}}\-- EHT-LTFs design considerations: EHT-LTFs reuse HE-LTFs in 20/40/80/160/80+80 MHz PPDU for \\  ~{} considering backward compatibility, while EHT-LTFs in 320/160+160/240/160+80 MHz PPDU pay more \\  ~{} attention to EHT-LTFs design methods with low peak to average power ratio and low overhead {[}20{]}\end{tabular}}                                                                                                      \\ \cline{2-2}
{\color{black} }                                               & {\color{black}\-- Proposals for forward compatibility as a requirement for IEEE 802.11be preamble {[}21{]}}                                                                                                                                                                                                                                                         \\ \cline{2-2}
{\color{black}}                                               & {\color{black}\-- Extending IEEE 802.11ax preamble puncturing patterns up to 240/320 MHz {[}22{]}}                                                                                                                                                                                                                                                                  \\ \cline{2-2}
\multirow{-7}{*}{{\color{black} \textbf{EHT Preamble Design}}} & {\color{black} \begin{tabular}[c]{@{}l@{}}\-- Simulation of the channel utilization gain when using a more effective channel puncturing than is used\\ ~{} in IEEE 802.11ax {[}23{]}\end{tabular}}                                                                                                                                                                      \\ \hline
{\color{black}\textbf{\begin{tabular}[c]{@{}c@{}}Higher-Order \\ Modulation Schemes\end{tabular}}}       & {\color{black} \begin{tabular}[c]{@{}l@{}}\-- Feasibility analysis of 4096-QAM in certain configurations, including using transmitting beamforming, \\ ~{} low number of streams and strict receiving EVM requirement or multiple receiving antennas {[}24{]}\end{tabular}}                                                                                           \\ \hline
\end{tabular}
\end{table*}
\subsection{New Bandwidth Mode}
In Fig. 3, the maximum obtainable transmission bandwidth over 2.4 GHz and 5 GHz frequency bands are 40 MHz consisting of two continuous 20 MHz and 160 MHz consisting of two continuous/discontinuous 80 MHz \cite{ref7}, respectively, which may not meet the requirements of high-throughput and low-latency services, such as 4k/8k video, AR or VR and online gaming. Currently, the new additional 6 GHz band (5.925 GHz-7.125 GHz, in U.S.) \cite{ref10}, with a total available bandwidth of 1.2 GHz is now under regulatory discussion for opening up to WLANs. The new features of 6 GHz band, such as up to 320 MHz bandwidth, will help to achieve the target of EHT: a maximum throughput of at least 30 Gbps. The 320 MHz bandwidth may be contiguous and located in the same 6 GHz band or noncontiguous and located in different bands (e.g., partly at 5 GHz band and partly at 6 GHz band). Following the existing bandwidth extension rule in WLAN, the 320 MHz bandwidth can be decomposed into two discontinuous 160 MHz bandwidths locating in the 5 GHz and 6 GHz bands, respectively.
\begin{table*}[!htbp]
\arrayrulecolor{black}
\renewcommand\arraystretch{1.2}
\centering
\caption{Applicable Multi-RU Combinations for Different Bandwidth Modes in EHT.}
\begin{tabular}{|c|l|l|}
\hline
\multicolumn{1}{|c|}{\textbf{Type}} & \multicolumn{1}{c|}{\textbf{Definition}} & \multicolumn{1}{c|}{\textbf{Allowed Combinations}}
\\ \hline
Small-size RU & 26-tone, 52-tone, 106-tone                                                                                       & \begin{tabular}[c]{@{}l@{}}- 26-tone RU + 106-tone RU for 20/40 MHz\\ - 26-tone RU + 52-tone RU for 20/40/80 MHz\end{tabular}                                                                                                                                                                                    \\ \hline
Large-size RU & \begin{tabular}[c]{@{}l@{}}242-tone, 484-tone, 996-tone, \\ 2 $\times$ 996-tone, 3 $\times$ 996-tone (new \\additional)\end{tabular} & \begin{tabular}[c]{@{}l@{}}- 242-tone RU + 484-tone RU for 80 MHz\\ - 484-tone RU + 996-tone RU for 160 MHz,\\ {~}{~}242-tone RU + 484-tone RU + 996-tone RU for 160 MHz\\ - 484-tone RU + 2 $\times$ 996-tone RU for 240 MHz, 2 $\times$ 996-tone RU for 240 MHz  \\ - 484-tone RU + 3 $\times$ 996-tone RU for 320 MHz, 3 $\times$ 996-tone RU for 320 MHz\end{tabular} \\ \hline
\end{tabular}
\end{table*}

At present, the EHT task group is discussing efficient approaches to utilize the contiguous and non-contiguous bandwidth. Parket \emph{et al.} \cite{ref11} and Wu \emph{et al.} \cite{ref12} proposed a flexible bandwidth extension strategy to obtain a wide bandwidth through multi-channel aggregation across 2.4 GHz, 5 GHz and 6 GHz bands, e.g., 20/40 MHz(2.4 GHz)+20/40/80/160 MHz(5 GHz)+80/160/320 MHz(6 GHz). In the early discussion, it has been agreed that only contiguous 240 MHz, noncontiguous 160+80 MHz, contiguous 320 MHz and noncontiguous 160+160 MHz, are supported as new bandwidth modes for EHT. Other noncontiguous bandwidth modes (e.g., 20+40+80 MHz) are inadvisable from the perspective of hardware design complexity. The new 240 MHz/160+80 MHz mode is constructed from three 80 MHz channels while the tone plan for each 80 MHz segment is the same as 80 MHz in IEEE 802.11ax. However, more discussions are still needed, such as whether it is formed by 80 MHz bandwidth puncturing of 320 MHz/160+160 MHz. For the new 320 MHz/160+160 MHz bandwidth, EHT should support the duplicated IEEE 802.11ax 160 MHz tone plan for the OFDMA tone plan. Since the preamble design for EHT is pending, the tone plan for non-OFDMA 320 MHz/160+160 MHz is still under discussion. For the OFDMA transmission in 320 MHz/160+160 MHz, combinations of large size RU (e.g., 2 $\times$ 996-tone RU+484-tone RU) are not determined up to now. The nature of non-OFDMA PPDU is the implementation of preamble puncturing for SU under the OFDMA format, while all RUs are assigned to the same user. For the existing 20/40/80/160/80+80 MHz bandwidth, EHT can reuse IEEE 802.11ax tone plans. It is worth noting that the data portion of the EHT PPDU uses the same subcarrier interval as that of IEEE 802.11ax \cite{ref13}.

\subsection{Multi-RU Support}
In IEEE 802.11ax, the RUs defined for DL and UL transmission are as follows: 26-tone RU, 52-tone RU, 106-tone RU, 242-tone RU, 484-tone RU, 996-tone RU and 2 $\times$ 996-tone RU. To enhance the spectral efficiency, the motion of multiple RUs to be assigned to a single user and new 3 $\times$ 996-tone RU have been approved in the EHT task group. EHT has some preliminary contributions in dealing with multi-RU combination schemes, coding and interleaving schemes, and multi-RU signaling designs. To achieve the trade-off between combination complexity and spectral efficiency, it is allowed to use some limited multi-RU combinations for the case with the bandwidth less or equal to 160 MHz, that is, small-size RUs (less than 242 tones) can only be combined with small-size RUs and large-size RUs (more than or equal to 242 tones) can only be combined with large-size RUs, but the mixture of small-size RUs and large-size RUs is unallowed \cite{ref4}. Table II lists the applicable multi-RU combinations for different bandwidth modes in EHT, where the combination of small-size RUs shall not cross 20 MHz channel boundary, and the combination of 26-tone RU and 52-tone RU for 20/40/80 MHz PPDU format and the combination of 26-tone RU and 106-tone RU for 20/40 MHz PPDU format are permitted. In terms of large-size RUs, the allowed large-size RU combinations are 242-tone RU + 484-tone RU for 80 MHz OFDMA/non-OFDMA PPDU format and 484-tone RU + 996-tone RU for 160 MHz OFDMA/non-OFDMA PPDU format. In IEEE 802.11ax, OFDMA only supports 2/4/8/16 users, while in EHT the multi-RU support could allow more flexible support for other values of the number of users, such as 5 or 6 users, and allowing up to 3 RUs to be assigned to a single user was proposed in \cite{ref14}. However, till now, the EHT task group has not reached a consensus on the maximum number of RUs assigned to a single user.\par
In terms of the data transmission in multiple RUs, the same or different set of transmission parameters, such as modulation and coding schemes (MCSs), interleaving schemes and the number of space-time streams, may be applied to combined RUs assigned to a user. There are four approaches to transmit the data in combined RUs \cite{ref15}: (1) all RUs are encoded and interleaved independently, (2) multiple RUs are encoded together, but each RU is interleaved independently, (3) multiple RUs need interleaver across RUs regardless of encoding, and (4) multiple RUs act as one logic/continuous RU. However, these potential approaches need to be analyzed and further evaluated under the constraints of hardware complexity and signaling overhead.\par
In addition to RU sizes/multi-RU combinations/multi-RU transmission, the EHT task group needs to make more efforts in signaling support for multi-RU PPDU, such as how to reuse/optimize the existing signaling methods (e.g., Bandwidth field, Allocation field or User field) in IEEE 802.11ax to indicate size/tone-mapping/combination of multi-RU. Different from IEEE 802.11ax where a unique STA-ID is used for one RU assigned to an STA, a matching STA-ID in EHT needs to be envisioned for the use of multi-RU.
\begin{figure*}[!htbp]
  \begin{center}
    \setlength{\abovecaptionskip}{-0.2cm}
    \scalebox{0.8}[0.8]{\includegraphics{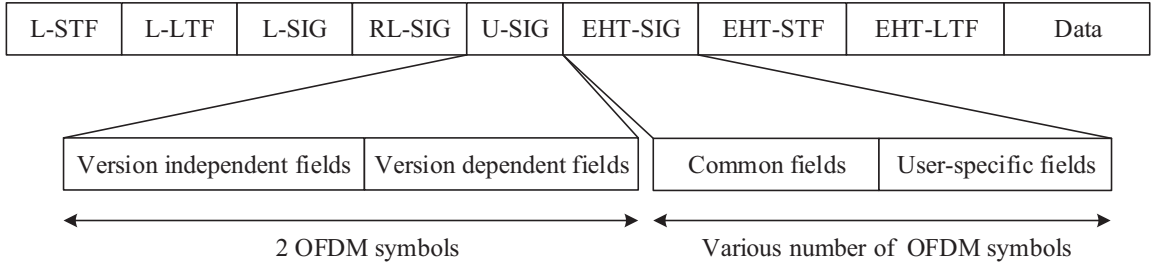}}
    \renewcommand{\figurename}{Fig.}
    {\color{black}\caption{The EHT PPDU format. For the backward-compatibility, the EHT frame format starts with the legacy field (i.e., the L-STF field, the L-LTF field, the L-SIG field, and RL-SIG field) used for frame detection, synchronization, carrying necessary information (e.g., MCSs and frame length). In the U-SIG field, the version independent fields include the PHY version identifier, UL/DL flag, BSS color, TXOP duration, etc. The version dependent fields in the U-SIG field consist of the similar information included in HE-SIG-A and other information for new EHT features. The common field in the EHT-SIG field contains information about multi-RU allocation, MCS, coding, etc. The user-specific fields in the EHT-SIG field carry individual dedicated information for multiple users. The EHT-STF field and the EHT-LTF field are used for channel estimation in the MIMO/OFDMA transmission.}}
  \end{center}
\end{figure*}
\subsection{EHT Preamble Design}
Observing the WLAN development process, each WLAN standard has its specific preamble, which provides functions including synchronization, channel estimation, auto-detection and necessary signaling, etc. Similar to IEEE 802.11ax, to support different technologies and scenarios, EHT should define at least a new preamble format for possible PPDU formats, such as EHT SU PPDU, EHT Trigger-based PPDU, EHT ER (extended range) SU PPDU and EHT MU PPDU. \par
As shown in Fig. 4, an EHT PPDU consists of a legacy part field (namely non-HT Short Training field (L-STF), legacy LTF field (L-LTF), legacy SIG field (L-SIG) and repeat legacy SIGNAL field (RL-SIG)), a universal SIG (U-SIG) field, an EHT-SIG field, an EHT Short Training field (EHT-STF), an EHT Long Training field (EHT-LTF) and a Data field \cite{ref4}. Specifically, to keep backward compatibility with legacy PPDUs operating in 2.4 GHz, 5 GHz, and 6 GHz bands, the legacy part field should be applied to the beginning of the EHT PPDU, which is used for frame detection, synchronization and carrying the necessary indication information (e.g., MCSs and frame length). For a PPDU with a bandwidth of 160 MHz or less, the legacy part is duplicated and can reuse the existing tone rotation \cite{ref16}. However, for a PPDU with a bandwidth wider than 160 MHz, the tone rotation is still not determined. To spoof IEEE 802.11ax devices and respect the length in the L-SIG field, the first symbol after L-SIG should be BPSK modulated in an EHT PPDU \cite{ref17}. To improve the robustness of L-SIG in outdoor scenarios and identify the EHT PPDU through automatic detection, the RL-SIG field is necessary and should be different from the RL-SIG field in the IEEE 802.11ax PPDU \cite{ref18}. \par Following the RL-SIG field, the EHT PPDU includes a two-OFDM-symbol U-SIG field like the HE SIGNAL A (HE-SIG-A) field in IEEE 802.11ax, which is used to carry the necessary information for the interpretation of EHT PPDUs \cite{ref4}. The U-SIG field contains both version independent fields and version  dependent fields. The version independent fields can be composed including PHY version identifier, UL/DL flag, BSS color, PPDU type, MCS, bandwidth, transmission opportunity (TXOP), etc. The version dependent fields likely consist of the similar information included in HE-SIG-A except the information included in the version independent fields as well as new information fields, such as the guard interval duration, EHT-STF/LTF size, space-time block coding flag, etc. Since discussions on other candidate characteristics for EHT are also underway, such as multi-link aggregation and multi-AP coordination, other PPDU-related descriptions and the number of bits for the U-SIG field are also pending. \par
To provide effective signaling support for an EHT PPDU sent to multiple users, such as the OFDMA and DL MU-MIMO resource allocation information, there should be a variable MCS and variable-length EHT-SIG field (immediately after the U-SIG) in an EHT PPDU. The EHT-SIG field consists of common fields and zero or several user-specific fields. The common field contains information about RU allocation, coding, MCS, the number of space-time streams, the duration of the guard interval, etc. The user-specific fields carry dedicated information for individual users. For the SU PPDU, the EHT-SIG field is composed of only the common field part without user-specific fields. In the EHT Trigger-based PPDU, since we can contain all needed information in the U-SIG field, the EHT-SIG is omitted. Similar with IEEE 802.11ax, for range extension, the sizes of the U-SIG field and EHT-SIG field in the EHT ER SU PPDU will likely be twice as the U-SIG field and EHT-SIG field in the EHT SU PPDU format, respectively. \par
The EHT-STF field and EHT-LTF field, as the last field of the EHT preamble, provides a field for users to estimate the MIMO channel, and EHT could support three types of EHT-LTF, including 1x EHT-LTF, 2x EHT-LTF and 4x EHT-LTF \cite{ref19}\cite{ref20}. Besides, in \cite{ref21}, reusing HE-LTFs for EHT-LTFs in 20/40/80/160/80+80 MHz EHT PPDU was recommended and the design methods for EHT-LTFs in 240/160+80/320/160+160 MHz EHT PPDU was proposed.\par
In the earlier preamble designs discussions, besides the backward compatibility with legacy PPDUs, the forward compatibility with EHT PPDUs is also raised as another issue to address \cite{ref5}. To solve the problem of increasingly complex preamble formats, it is important to minimize design burden and limit the complexity while keeping forward compatibility with future IEEE 802.11 generations.\par
Preamble puncturing is an effective approach to enhance the channel utilization and improve the transmission rate. With such bandwidth wider than 160 MHz in EHT, preamble puncturing will demand more complicated hardware operations and more flexible preamble puncturing patterns \cite{ref22}\cite{ref23}, e.g., extending IEEE 802.11ax preamble puncturing patterns up to 240/320 MHz or applying puncturing to primary channels to increase channel access opportunities.
\subsection{Higher-Order Modulation Schemes}
To further enhance the peak rate, compared with IEEE 802.11ax whose highest-order modulation scheme is 1024-QAM, a higher-order modulation scheme, i.e., 4096-QAM, has been suggested for EHT, where one modulation symbol can carry 12 bits. Theoretically, given the same coding rate, EHT can achieve 20\% higher transmission rate compared with IEEE 802.11ax, thereby enabling its users to obtain higher transmission efficiency while requiring higher EVM. Preliminary simulation results in \cite{ref24} reveal that applying 4096-QAM is feasible in certain configurations, such as using transmitting beamforming, low number of streams, strict EVM requirement or multiple receiving antennas. Other feasible configurations, such as coding strategies, also need further explorations and verifications through simulations and experiments. Besides, EHT-MCSs should be respectively defined for both SU transmission and MU transmission.\par
To improve the received signal quality, i.e., signal-noise ratio (SNR), as well as the transmission robustness, EHT may still support dual-carrier modulation (DCM), which enables the same information to be modulated over a pair of subcarriers. In IEEE 802.11ax, DCM is only applicable to the MCS 0/1/3/4 and 1/2 spatial streams to satisfy the high-reliability requirements \cite{ref3}. With the support of other candidate features in EHT, such as multi-AP support for transmitting the same frame to a user or HARQ support, DCM may be applicable to higher-order modulation schemes (e.g., MCS 5/6) or more spatial streams (e.g., 3/4 spatial streams) to guarantee the high transmission reliability.
\begin{figure*}[!t]
  \begin{center}
    \setlength{\abovecaptionskip}{-0.3cm}
    \scalebox{0.4}[0.4]{\includegraphics{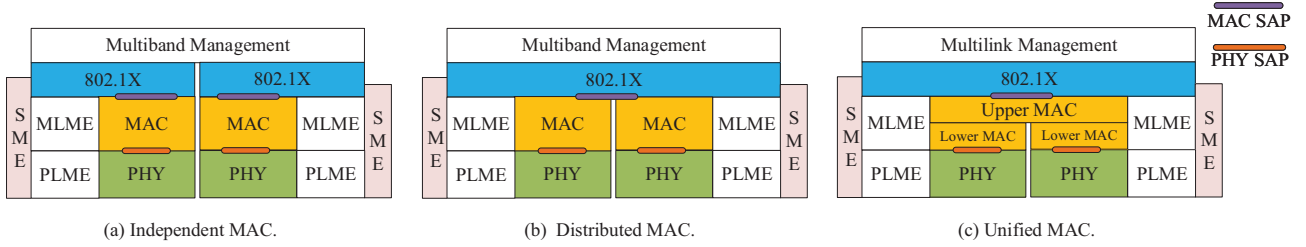}}
    \renewcommand{\figurename}{Fig.}
    \caption{Two existing MAC architectures for multi-band and an enhanced MAC Architecture for multi-link aggregation. (a) Independent MAC for multi-band. (b) Distributed MAC for multi-band. (c) Unified MAC for multi-link.}
  \end{center}
\end{figure*}

\begin{table*}[!htbp]
\arrayrulecolor{black}
\renewcommand\arraystretch{1.2}
\centering
{\color{black}\caption{Summary of Multi-link Operations and Related Works.}}
\begin{tabular}{|c|l|}
\hline
\multicolumn{1}{|l|}{{\color{black} }}                                                                   & \multicolumn{1}{c|}{\textbf{{\color{black} Contributions}}}                                                                                                                                                                                                                                                                  \\ \hline
{\color{black} }                                                                                         & {\color{black} \begin{tabular}[c]{@{}l@{}}\-- A unified framework design that addresses the key use cases (load balancing and aggregation) and\\ ~{} keeps within the current 802.11 architecture and definition {[}8{]}\end{tabular}}                                                                                  \\ \cline{2-2}
\multirow{-2}{*}{{\color{black} \textbf{\begin{tabular}[c]{@{}c@{}}Multi-link MAC \\ Architecture\end{tabular}}}} & {\color{black}\-- Multi-Link reference modes and a potential MAC protocol model designs {[}25{[}26{]}}                                                                                                                                                                                                                \\ \hline
{\color{black}}                                                                                         & {\color{black}\-- CCA indication and per-20 MHz bitmap for multi-link, channel access in the primary channel set {[}22{]}}                                                                                                                                                                                            \\ \cline{2-2}
{\color{black} }                                                                                         & {\color{black} \begin{tabular}[c]{@{}l@{}}\-- Channel access based on one primary channel to reduce scanning latency and energy consumption for \\ ~{} 6 GHz operations {[}27{]}{[}28{]}\end{tabular}}                                                                                                                      \\ \cline{2-2}
{\color{black}}                                                                                         & {\color{black}\-- Channel access with a temporary primary channel only when the primary channel is not available {[}22{]}}                                                                                                                                                                                            \\ \cline{2-2}
{\color{black}}                                                                                         & {\color{black} \begin{tabular}[c]{@{}l@{}}\-- Simulation analysis of the average area throughput and channel utilization by using \\ ~{} channel access with a temporary primary channel when the primary channel is busy {[}29{]}-{[}31{]}\end{tabular}}                                                                    \\ \cline{2-2}
\multirow{-5}{*}{{\color{black}\textbf{ \begin{tabular}[c]{@{}c@{}}Multi-link \\ Channel Access\end{tabular}}}}   & {\color{black} \begin{tabular}[c]{@{}l@{}}\-- Independent channel access with multiple primary channels to improve spectral  efficiency and ensure \\ ~{} backward compatibility {[}33{]}{[}34{]}\end{tabular}}                                                                                                             \\ \hline
& {\color{black} \begin{tabular}[c]{@{}l@{}}\-- Fast switching between multiple links to reach high spectrum utilization and load balancing, an\\ ~{} architecture which provides a unified solution for control/data separation to reduce transmission delay \\ ~{} and improve spectrum utilization {[}35{]}\end{tabular}} \\ \cline{2-2}
{\color{black} }                                                                                         & {\color{black}\begin{tabular}[c]{@{}l@{}}\-- Control information (e.g., control frames, A-Control fields) for a channel transmitted in a different \\ ~{} channel/band {[}28{]}\end{tabular}}                                                                                                                             \\ \cline{2-2}
{\color{black} }                                                                                         & {\color{black}\begin{tabular}[c]{@{}l@{}}\-- A control plane and data plane decoupled WLAN architecture design for efficient resource management \\~{} and high reliability {[}37{]}\end{tabular}}                                                                                                                    \\ \cline{2-2}
{\color{black}}                                                                                         & {\color{black}\-- Channel access and data transmission in both of asynchronous mode and synchronous mode {[}38{]}{[}39{]}{[}41{]}}                                                                                                                                                                                    \\ \cline{2-2}
{\color{black}}                                                                                         & {\color{black}\begin{tabular}[c]{@{}l@{}}\-- A mixed multiple links system design to support mixed asynchronous and synchronous transmission \\ ~{} mode {[}40{]}\end{tabular}}                                                                                                                                        \\ \cline{2-2}
\multirow{-6}{*}{{\color{black} \textbf{\begin{tabular}[c]{@{}c@{}}Multi-link\\ Transmission\end{tabular}}} }
&{\color{black}\begin{tabular}[c]{@{}l@{}}\-- Simulation of synchronous and asynchronous multi-band/multi-channel operations to achieve significant \\ ~{} performance enhancements under heavy interference {[}27{]}\end{tabular}}                                                                                        \\ \hline
\end{tabular}
\end{table*}
\subsection{Summary of the PHY Enhancements for EHT}
To enable the new PHY and MAC modes to support a maximum throughput of at least 30 Gbps, EHT introduces several PHY enhancement technologies to accomplish this goal, including new bandwidth modes, multi-RU support, enhanced preamble designs and 4096-QAM. From the literatures surveyed in this section, we can observe that the EHT task group has some preliminary contributions in the designs and verifications of the enhanced PHY technologies. However, since the EHT standardization process has just started and many things are open, the research work on these enhanced PHY technologies still needs to be continued, such as tone plan designs for new bandwidth modes, coding and interleaving schemes for multiple RUs assigned to a single user, tone-mapping for multi-RU, multi-RU combination schemes and signaling designs, feasible configurations for applying 4096-QAM, preamble designs. In the standardization process of EHT, the EHT task group can refer to the PHY designs of the latest IEEE 802.11ax (e.g., backward compatibility and preamble designs of the TB PPDU and ER PPDU), and then develop the PHY technical specifications (e.g., introducing more efficient encoding/decoding methods than Low Density Parity Check (LDPC)).\par
In the WLAN specifications, the PHY provides an interface to the MAC layer through an extension of the generic PHY service interface, such as TXVECTOR, RXVECTOR. Specifically, the MAC layer uses the TXVECTOR to provide the PHY with per-PPDU transmission parameters, and the PHY uses the RXVECTOR to inform the MAC layer of the received PPDU parameters. In the process of developing the EHT technical specifications, EHT also needs to clearly define the interface parameter conditions and corresponding values, such as the number of spatial streams, MCS, multi-RU allocation and channel width of the PPDU.
\par
In the following Section III-VII, we will mainly focus on the MAC enhancements and cross-layer techniques between PHY and MAC layers of EHT.

\section{Multi-link Operations}
The simultaneous operating over 2.4 GHZ, 5 GHz and 6 GHz bands is a noteworthy feature of the EHT. To achieve highly efficient utilization of all available spectrum resources, new spectrum management, coordination and transmission mechanisms over the 2.4 GHz, 5 GHz and 6 GHz, are urgently needed. In this section, we will overview some promising multi-link aggregation technologies summarised in Table III, including the enhanced multi-link MAC architecture, the multi-link channel access and the multi-link transmission.
\begin{figure*}[bp]
  \begin{center}
    \setlength{\abovecaptionskip}{-0.2cm}
    \scalebox{0.8}[0.8]{\includegraphics{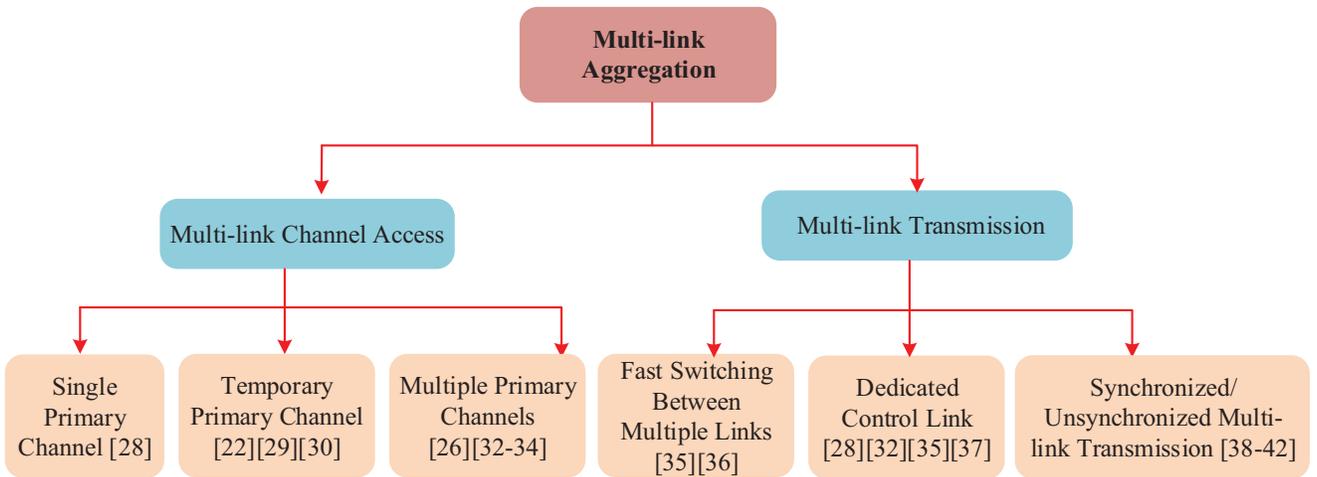}}
    \renewcommand{\figurename}{Fig.}
    {\color{black}\caption{Overview of multi-link operations.}}
  \end{center}
\end{figure*}
\subsection{Enhanced MAC Architecture for Multi-link Aggregation}
As described in PAR \cite{ref5}, one of the goals of the EHT TG is to meet high-throughput and stringent real-time delay requirements of 4k/8k video, VR or AR, online gaming, etc. Therefore, the multi-link operation to meet those PAR requirements becomes a hot topic being discussed by the EHT task group. There is a trend that STAs are moving into dual-band/tri-band parallel architectures aggregated across 2.4 GHz, 5 GHz and 6 GHz bands, which requires new management specifications and usage rules for multiple bands. The current IEEE 802.11 protocol recommends two multi-band MAC architectures \cite{ref7} to provide different technical support for multi-band operations, i.e., Independent MAC (nontransparent FST) and Distributed MAC (transparent FST) as shown in Fig. 5. Both architectures can provide a renegotiation pipe for seamless session transfer from one channel to another in the same or different frequency bands. Nevertheless, there exists a limitation that MSDUs belonging to single TID can only use a single band and/or channel, resulting in significant MAC overheads for renegotiations. For example, when switching sessions through legacy FST, STAs need to initiate the FST Setup Request and Response frame exchanges as well as the FST Ack Request and Response frame exchanges, resulting in significant MAC overhead for renegotiation. To eliminate the need for various management/data plane renegotiations for fast session transfer, an essential MAC improvement of the new MAC architecture (named Unified MAC for multi-link) \cite{ref8}\cite{ref25}\cite{ref26} in Fig. 5 should be that MSDUs belonging to single TID or different TIDs can be transmitted over multiple bands and/or channels concurrently or non-concurrently. In all three of the above MAC architectures, how and in which architectures for other multi-band operations will be supported also need to be better understood. \par
{\color{black}In the Independent MAC architecture \cite{ref7}, different MAC service access points (SAPs) are presented to upper layers since different MAC addresses are used before and following an FST, and upper layers are responsible for managing the session transfer to balance traffic load between different bands/channels. Since the function of the multi-band management entity is restricted to coordinating the setup and teardown of an FST with no access to other local information of station management entities (SMEs), local information of an SME, such as robust security network associations (RSNAs), security keys and packet number (PN) counters, needs to be re-established for the new band/channel. In the Distributed MAC architectures \cite{ref7}, only one MAC SAP that is identified by the same MAC address is presented to the higher layers, making upper layers unaware of the session transfer between different bands/channels. The local information of each SME can be shared between multiple bands/channels, including block ack (BA) agreements, traffic streams, association state, RSNA, security keys and PN counters. In the Unified MAC architecture \cite{ref25}, it contains only one MAC sublayer management entity and one MAC sublayer. Unlike the Independent MAC and Distributed MAC, the MAC protocol stack is divided into Upper MAC which supports most MAC operations (e.g., A-MSDU aggregation/de-aggregation, sequence/packet number assignment, encryption/decryption and integrity protection/check, and fragmentation/defragmentation) and Lower MAC which supports a small number of MAC operations (e.g., MPDU header and cyclic redundancy check (CRC) creation/validation and MPDU aggregation/de-aggregation). Besides, the Unified MAC can support the dynamic transfer of a TID among multiple links. The traffic is put into the queue and uses all or part of the available channels for the concurrent or non-concurrent transmission.}

\subsection{Multi-link Operation over Wideband and Noncontiguous Spectra}
By utilizing multi-link aggregation across 2.4 GHz, 5 GHz and 6 GHz bands, a multi-link capable device can parallelly transmit frames through multiple links, thereby achieving higher throughput and improving the network flexibility compared with IEEE 802.11ax. However, considering the existing legacy devices in 2.4 GHz, 5 GHz and 6 GHz, available links may be restricted. Thus, how to access in multi-link and to transmit frames over multiple links which may be beneficial to wideband and noncontiguous spectrum availability need further studies. As shown in Fig. 6, several promising multi-link channel access methods and multi-link transmission modes have been proposed in the EHT task group, including the channel access based on one primary channel, the channel access based on multiple primary channels, the dedicated control link, the fast link switching and the synchronized/unsynchronized multi-link transmission.\begin{figure*}[!hb]
  \begin{center}
    \setlength{\abovecaptionskip}{-0.2cm}
    \scalebox{0.46}[0.46]{\includegraphics{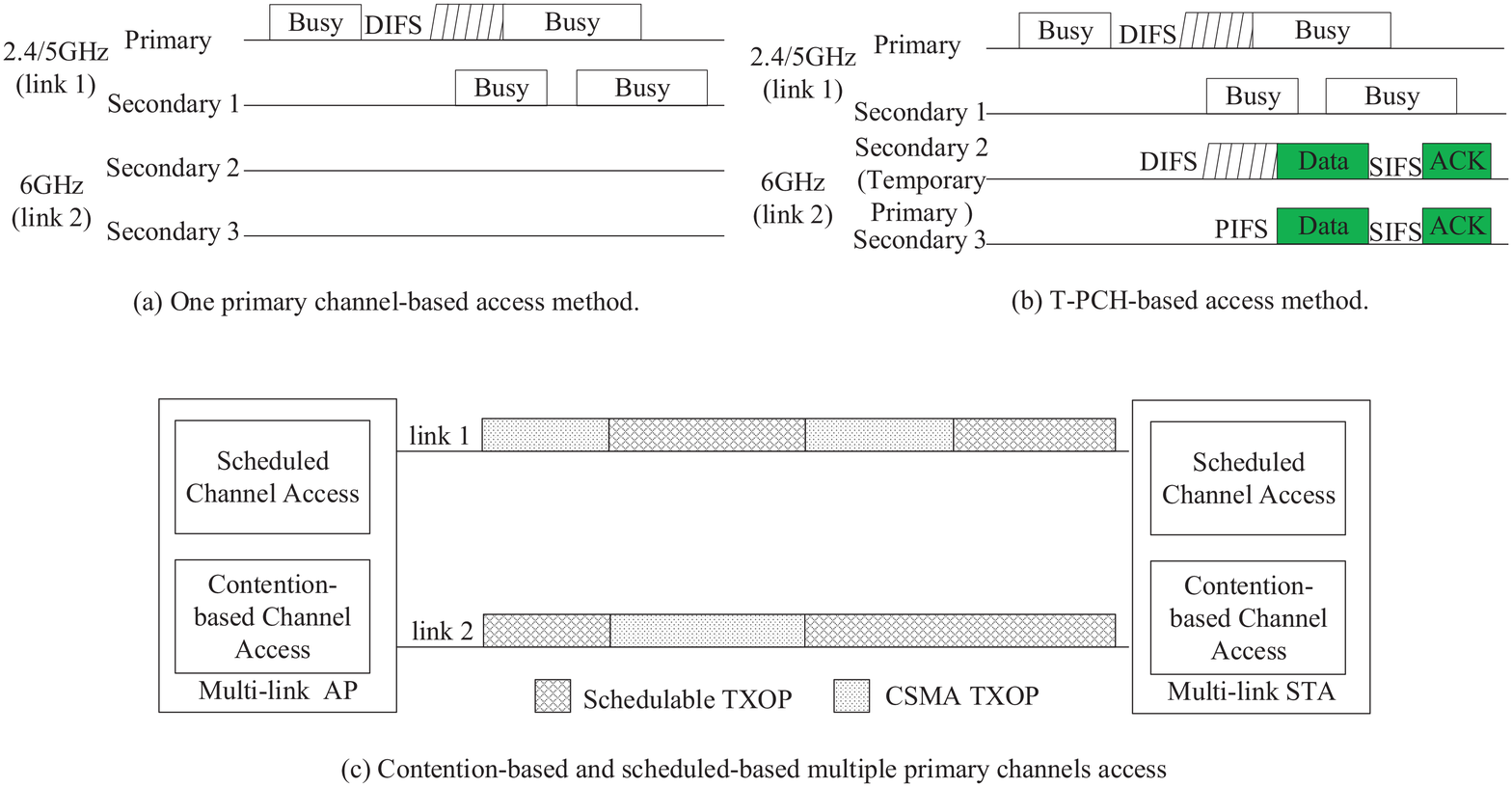}}
    \renewcommand{\figurename}{Fig.}
    \caption{Channel access methods for multiple links. (a) As the legacy channel access operation, the contention-based channel access for multi-link is performed in one primary channel. (b) If and only if the primary channel is unavailable, the contention-based channel access is performed on the temporary primary channel set on the secondary channel, otherwise it is performed on the primary channel. (c) The AP/STA simultaneously run the scheduled-based module and the contention-based module on different links. The scheduled-based module is responsible for scheduling MSDUs for real-time applications during the schedulable TXOPs, and the contention-based module is responsible for transmitting MSDUs from non-real-time applications during the TXOPs.}
  \end{center}
\end{figure*}
\subsubsection{Multi-link Channel Access}
In current WLANs, channel access mechanisms (e.g., clear channel assessment (CCA) indication and per-20MHz bitmap) are only defined for a single link at 20/40/80/160 MHz channel. However, future Wi-Fi devices are expected to be multi-link capable and have wider channel width at most 320 (160+160) MHz across 2.4 GHz, 5 GHz and 6 GHz bands. At present, these channel access mechanisms for multi-link have not yet been defined in EHT. Following the existing channel access logic, CCA indication and per-20 MHz bitmap will evolve from one link to multiple links\cite{ref22}. For example, in terms of 320 MHz bandwidth, CCA demands to add secondary 160 indications. and per-20 MHz bitmap demands more preamble puncturing patterns which may even be applied to the primary channel. In this article, we classify channel access methods for multi-link into two categories: channel access based on one primary channel and channel access based on multiple primary channels.
\paragraph{Channel access based on one primary channel}
As the legacy channel access operation, the multi-link channel access may be performed in one primary channel. In Fig. 7(a), the STA simultaneously accesses 2.4/5 GHz and 6 GHz bands by performing contention-based access at the primary channel, which could limit channel pollution caused by scanning in the 6 GHz band and reduce scanning latency and energy consumption for 6 GHz operations \cite{ref27,ref28}. However, at 2.4 GHz, 5 GHz and 6 GHz bands, there exist legacy STAs. Accordingly, obtaining TXOP at 6 GHz depends not only on the occupancy of the target primary channel at 2.4/5 GHz but also on the activity of the secondary channels at 6 GHz. Such a channel access method has less flexibility in channel selection and usage for multi-link, especially in dense deployment scenarios, which may significantly degrade the multi-link Wi-Fi system's performance due to the probe storming effect, thereby increase collisions and reduce access opportunities on the primary channel. For this reason, in addition to designing new preamble puncturing schemes \cite{ref22}, channel coordination between legacy devices and EHT devices through wired/wireless tunnels might be an excellent option, e.g., allowing EHT devices to restrict channel access of legacy STAs or enforce channel selection changes in legacy APs.
\paragraph{Channel access based on multiple primary channels}
In the current IEEE 802.11 protocols, since a device can obtain its TXOP on its primary channel, spectral resources are hardly utilized if congestion occurs in the primary channel. To overcome this problem, a temporary primary channel (T-PCH) \cite{ref22}\cite{ref29}\cite{ref30} can be set on the secondary channels to increase the channel use opportunities when the primary channel is unavailable. In Fig. 7(b), the STA can carry out carrier sensing on the T-PCH as well as on the primary channel, and it obtains TXOP on the T-PCH if the T-PCH is idle for a required duration when the primary channel is busy. After the T-PCH becomes busy while the primary channel becomes idle again, the STA can be allowed to return from the T-PCH to the primary channel immediately. By doing this, the STA may obtain more TXOP on the more idle channels. Through computer simulation \cite{ref31}, it is confirmed that the proposed T-PCH can improve the average area throughput and channel utilization. In a real environment, such a T-PCH operation could ensure that it does not affect the systems' performance of those legacy STAs already operated in these channels and not damage the fairness between new types of APs and other legacy APs severely.\par

Since the presence of T-PCH depends on the status of the primary channel, such channel access dramatically limits the use of the idle channels. An intuitive idea should be that the STA may perform channel access on multiple links independently. Each link performs specific functionality independently, e.g., Enhanced Distributed Channel Access (EDCA) and CCA. This method has higher backward compatibility compared to the legacy single-link reference architecture, which is difficult for the coordination of multi-link operations by the upper layers \cite{ref26}. Via simulations, the probability of successfully obtaining channels concurrently on two links is not high \cite{ref32}. Because the back-off of multi-link is finished at different times, it has different operation rules for different kinds of the multi-link transmission \cite{ref33}. For the independent multi-link transmission, the back-off in each link reuses the existing back-off rules. For the simultaneous multi-link transmission, the back-off procedure in multiple links may be as follows: \begin{figure*}[!bp]
  \begin{center}
    \setlength{\abovecaptionskip}{-0.2cm}
    \scalebox{1}[1]{\includegraphics{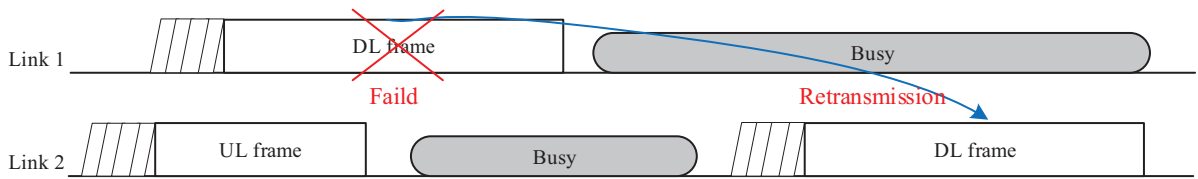}}
    \renewcommand{\figurename}{Fig.}
    \caption{An example of the multi-link transmission with collaboration. When the DL frame in the link 1 fails, it can be retransmitted immediately in the available link 2 to reduce the waiting latency.}
  \end{center}
\end{figure*}
\begin{figure*}[b]
  \begin{center}
    \setlength{\abovecaptionskip}{-0.2cm}
    \scalebox{1.1}[1.1]{\includegraphics{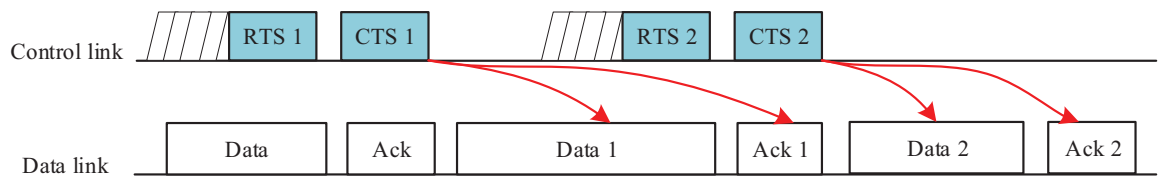}}
    \renewcommand{\figurename}{Fig.}
    \caption{An example of decoupling the data and control planes on different links, where the control link is only used to carry control and management frames for data exchange and the data link is used to data frame and the ack frame exchanges.}
  \end{center}
\end{figure*}(i) When the back-off counter in one link is reduced to 0 first, and aggregate other links, it will cause fairness issue in other links. (ii) When the back-off counter in all links are reduced to 0, and then transmit packets simultaneously on multi-link, the STA will have less channel access chance than the legacy STA. In this regard, dynamic bandwidth negotiation could be supported in multi-link.\par
However, different kinds of channel access methods (e.g., EDCA and triggered uplink channel access \cite{ref3}) and the diverse requirements of the current and future applications are not considered in the aforementioned channel access approaches. To satisfy different transmission requirements for different services, which are real-time and non-real-time applications, the optimized multiple primary channels access method was also proposed in \cite{ref34} as shown in Fig. 7(c), where an AP can simultaneously run two channel access function modules on different links, namely, the scheduled-based module and the contention-based module. The scheduled-based module is responsible for scheduling MSDUs from real-time applications during the schedulable TXOPs of multi-link, and the contention-based module is responsible for transmitting MSDUs from non-real-time applications mainly during the Carrier Sense Multiple Access (CSMA) TXOPs of multi-link. In this way, the backward compatibility and coexistence with legacy channel access methods would guarantee the multi-link capable STAs operating with different access methods operate in the same link or different links.

\subsubsection{Multi-link Transmission}
\paragraph{Fast switching between multiple links}
In general, the wider the transmission bandwidth, the higher the occurrence probability of co-channel and adjacent channel interference on neighboring nodes will be, degrading the spectrum efficiency. Thus, dynamic link switching based on wireless link states is a critical technology to reduce the strong interference from neighboring nodes. When the QoS of the current in-service link cannot meet the requirements, a multi-link capable AP/STA can switch the control/management frames and data to other idle and high-quality links. For the type of existing switching with negotiation in current IEEE 802.11 specifications \cite{ref7}, there is still significant MAC overhead related to multi-link operations. For example, when switching sessions between links, STAs usually require necessary frame exchanges as data from a single TID and corresponding BA/Ack can only be allocated to the same link. For the type of flexible and new switching without negotiation, data from single TID and corresponding BA/Ack frames should be transmitted in all links, and operating actions in one link should also be conducted to all other links, such as key negotiation, BA negotiation, and power-saving negotiation. Fast switching between multiple links requires devices to efficiently select channels in different links to reach high spectrum utilization. Seamless switching between different links helps address the use-case for efficient retransmission, load balancing and coexistence constraints \cite{ref35}. In Fig. 8, the DL frame failed in link 1 can be retransmitted in the available link 2 for reducing the waiting latency, and the channel diversity can smoothen out link fluctuations. Based on various load balancing methods presented in \cite{ref36} including admission control, association management, transmission range control, and association control, STAs can decide to switch all traffic or partial traffic from one overloaded link to another underloaded link to improve QoS. For example, based on types of traffic, the STA can transmit high-throughput and low-latency services on one link (e.g., 5/6 GHz) and transmit delay-insensitive services on another link (e.g., 2.4 GHz).
\begin{figure*}[!b]
  \begin{center}
    \setlength{\abovecaptionskip}{-0.2cm}
    \scalebox{0.9}[0.9]{\includegraphics{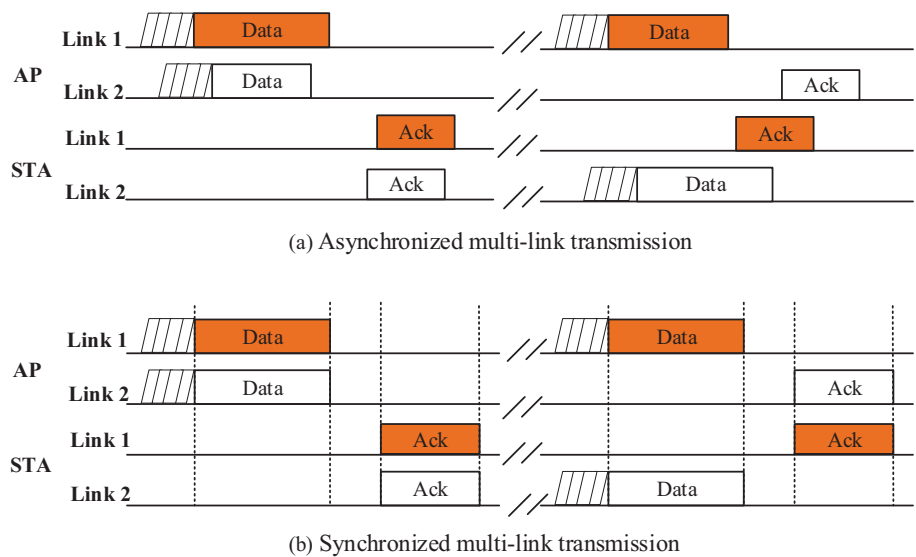}}
    \renewcommand{\figurename}{Fig.}
    \caption{An example of the asynchronized and synchronized multi-link transmissions. Regardless of transmitting and receiving simultaneously or transmitting and receiving non-simultaneously, the transmission starting time on multiple links under the asynchronized transmission mode is unaligned. The synchronized transmission with aligned transmission starting time allows a device to transmit and receive frames simultaneously/non-simultaneously on multiple links.}
  \end{center}
\end{figure*}

\begin{table*}[!t]
\centering
\renewcommand\arraystretch{1.2}
{\color{black}\caption{Summary of Synchronized/Unsynchronized Multi-link.}}
\begin{tabular}{|c|c|c|}
\arrayrulecolor{black}
\hline
\textbf{}                       & \textbf{{\color{black}Unsynchronized Multi-link}}                                                                                                                                                                                                                                                                      & {\color{black}\textbf{Synchronized Multi-link}  }                                                                                       \\ \hline
{\color{black}\textbf{Transmission Capability} }   & {\color{black}With simultaneous TX \& RX capability }                                                                                                                                                                                                                                                                  & {\color{black}Without simultaneous TX \& RX capability }                                                                                \\ \hline
{\color{black}\textbf{Power Leakage} }             & \begin{tabular}[c]{@{}c@{}}{\color{black}Sufficient frequency isolation or interference}\\ {\color{black}cancellation for concurrent DL and UL.}
\end{tabular}                                                                                                                                                                                             & \begin{tabular}[c]{@{}c@{}}{\color{black}Sufficient frequency isolation or interference}\\ {\color{black}cancellation for concurrent DL and UL.} \end{tabular}                                                                               \\ \hline
{\color{black}\textbf{Channel Access}  }           & \begin{tabular}[c]{@{}c@{}}{\color{black}Independent channel access in each link,}\\ {\color{black}minor or no change for standards}\end{tabular}                                                                                                                                                                                                                 & \begin{tabular}[c]{@{}c@{}}{\color{black}Dependent channel access for multi-links,}\\ {\color{black}complex channel access rules}\end{tabular}                                        \\ \hline
{\color{black}\textbf{Start Time of Transmission} }& {\color{black}Unaligned }                                                                                                                                                                                                                                                                                              & {\color{black}Aligned  }                                                                                                                \\ \hline
{\color{black}\textbf{End Time of Transmission}  } & {\color{black}Unaligned }                                                                                                                                                                                                                                                                                             & {\color{black}Aligned }                                                                                                                 \\ \hline
{\color{black}\textbf{PPDU Parameters in Each Link} } & {\color{black}Independent PPDU length, bandwidth, MCS, et al.                                                                                                                                                                                                                                                                                            }& {\color{black}Dependent PPDU length, bandwidth, MCS, et al. }                                                                                                                                                 \\ \hline
{\color{black}\textbf{Spectrum Utilization}  }     & {\color{black}High }                                                                                                                                                                                                                                                                                                   & {\color{black}Low }                                                                                                                     \\ \hline
{\color{black}\textbf{QoS Issues}}                & \begin{tabular}[c]{@{}c@{}}{\color{black}Non-sequential packet receptions due to difference of}\\ {\color{black}transmission timing and frame length between links, }\\ {\color{black}failed  reception due to leakage to adjacent channel, }\\ {\color{black}unnecessary retransmission due to difference of}\\ {\color{black}channel quality between links \cite{ref41}}\end{tabular} & \begin{tabular}[c]{@{}c@{}}{\color{black}Unnecessary retransmission due to difference of }\\{\color{black}channel quality between links \cite{ref41}}\end{tabular} \\ \hline
\end{tabular}
\end{table*}

\paragraph{Dedicated control link}
The legacy STA exchanges packets by utilizing numerous sequential control/management operations on the same channels only in one link (e.g., 2.4 GHz or 5 GHz). The data and control/management operated in separate time will lead to large transmission delay and low spectrum utilization. These issues could be mitigated by decoupling the data and control planes over different links \cite{ref35}. Decoupling the data and control planes allows updating regular control/management frames on one of the links, leaving the other links primarily for data exchange. In Fig. 9, the control link can arrange every communication over different data channels, which requires a method that the receiver knowing exactly which channels to receive the data through negotiation or intelligent algorithms. Also, other control information (e.g., control frames, MAC/PHY header) can be transmitted in a dedicated control link and allowing the out-of-link exchange of control information can reach more efficient resource allocation \cite{ref28}. The complete decoupling of the data and control planes should also split data packets into two parts: the data part and the control part transmitted over multiple links. However, since the data are transmitted over different links, the non-sequential order of data reception may happen due to the difference in transmission timing. Therefore, more researches are still needed to build a more robust and efficient multi-link system with decoupling the data and control planes. To solve the problems of inefficient management and poor reliability of the existing distributed WLAN, a new control plane and data plane decoupled WLAN architecture is proposed in \cite{ref37}, which is a centralized control network architecture with control plane and data plane decoupling, in which AC (Access Controller) controls and manages all the APs and STAs through the control plane in the low-frequency band, and the AP provides data transmissions for STAs through the data plane in the high-frequency band.

\paragraph{Synchronized/unsynchronized multi-link transmission}
In Fig. 10, two types of multi-link transmissions are illustrated according to the capability of simultaneous UL/DL transmissions on multiple links, namely, asynchronized multi-link and synchronized multi-link transmissions \cite{ref38,ref39,ref40,ref41,ref42}. Both types of multi-link operations have their pros and cons as listed in Table IV. {\color{black}For the asynchronized multi-link transmission, a device can transmit frames with unaligned transmission starting time on multiple links. Each link has independent channel access and its own primary channel as well as EDCA parameters. Using different channel conditions and regulatory power limits for different links can achieve optimal throughput with per link individual MCS. The synchronized multi-link transmission means that a device shall transmit frames on multiple links with aligned transmission starting time. Waiting for CCA idle on all links before the transmission will waste time on early idle links, and it may need schemes to hold the idle channel, e.g., controlling the maximum standby time based on the idle time prediction of the probabilistic neural network \cite{ref43}. }In both synchronized multi-link transmission and unsynchronized multi-link transmission, the non-simultaneously transmitting and receiving on one or more links are permitted, and transmitting on one link while simultaneously receiving on another link may be supported. In addition, more efforts ranging from the PHY/MAC protocol designs to the developments of theoretical foundations need to be made to realize feasible asynchronized/synchronized multi-link operations in EHT. The performance of asynchronized/synchronized multi-link operations in future various IEEE 802.11 networks, such as multi-AP collaboration networks, need to be further evaluated from the theoretical aspects, such as network capacity and achievable transmission rate.

\subsection{Summary of Multi-link Operations}
The multi-link operations are suggested as a key candidate feature of EHT to improve the peak throughput and reduce latency and jitter. In this section, the technical issues related to the efficient multi-link operations are investigated, including multi-link MAC architectures, channel access methods and transmissions over multi-link.\par
The extremely efficient channel access and transmission designs call for new management functions in the enhanced multi-link MAC architecture, including multi-link setups (e.g., multi-link association), teardown operations of an existing multi-link setup agreement, multi-link MAC address managements, BA/Ack sessions and security managements in multiple links, etc.\par
In current WLAN specifications, the channel access is designed for a single link, how to access over the multi-link needs to be carefully discussed in EHT. The simultaneous multi-link channel access may be performed with one or more primary channels. Since the link status (busy/idle) of each link may be different and the per-link back-off procedure is performed, the channel access operation with multiple primary channels is more complicated than the channel access operation with a single primary channel. In this case where a multi-link capable STA performs a contention independently on each link, to align TXOPs across multiple links, new back-off mechanisms and medium reservation processes need to be discussed carefully, such as enforcing the back-off counters to reduce to zero, pausing the back-off procedure or resetting the back-off counters of all links to the same value. Besides, the method of the link status determination needs more consideration, e.g., energy detection only on multi-link, packet detection only on multi-link, or energy detection combined with packet detection on multi-link.\par
This section also shows how to use different multi-link transmission options to support different application use cases. The multi-link transmissions can be classified into the fast link switching for coexistence constraints/load balancing, the control/data separation for efficient channel utilization, independent transmission and simultaneous transmission. For each multi-link transmission option, this section does not cover the detailed discussions of the BA/ack agreements. Theoretically, compared to the existing single-link transmission mode, the multi-link transmissions can double link capacity at the same time resource. However, in the real world, performance gains of the multi-link transmissions may be hindered by legacy single-link devices. Therefore, designing the effective multi-link transmission schemes need to take into account the spectrum utilization, the design complexity, and the activities of legacy devices operating at the same link(s). Moreover, the simultaneous transmitting and receiving operation over the multi-link can cause inter-link interference due to power leakage unless their links are set a minimum separation or sufficiently far. Since the large guard separation between adjacent links can reduce the spectrum utilization, we need to explore some advanced analog/digital interference cancellation/suppression schemes for multi-link transmissions.

\begin{table*}[!htbp]
\renewcommand\arraystretch{1.9}
\centering
\caption{Enhanced Feedback Reduction Schemes.}
\begin{tabular}{|c|c|l|l|}
\hline
\textbf{Scheme} & \textbf{Method} & \textbf{\begin{tabular}[c]{c@{}c@{}}Pros\end{tabular}} & \textbf{\begin{tabular}[c]{c@{}c@{}}Cons\end{tabular}}\\        
\hline
\multirow{4}{*}{\begin{tabular}[c]{@{}c@{}}\rotatebox{90}{\textbf{Enhanced Schemes}}\end{tabular}}
& \emph{$\phi$} Only Feedback {\color{black}[48][49]}& \makecell[l]{Existing in IEEE 802.11ah and with minor\\ design efforts.} & Only single data stream is supported.\\
\cline{2-4}
& Time Domain Channel Feedback {\color{black}[47]}& \makecell[l]{Existing in IEEE 802.11ad/ay and with \\minor design efforts.} & \makecell[l]{May need additional signaling to identify \\tappositions and the extra matrix.}\\
\cline{2-4}
& Differential Given's Rotation {\color{black}[47][50]}& \makecell[l]{Improving from IEEE 802.11ax, IEEE \\802.11ay and with minor design efforts.} & \makecell[l]{May need additional processing and have \\error propagation.}\\
\cline{2-4}
& Variable Angle Quantization {\color{black}[48][49]}& \makecell[l]{Improving from IEEE 802.11ax and with \\minor design efforts.} & \makecell[l]{May need additional processing and \\signaling to indicate quantization levels.}\\
\hline
\multirow{4}{*}{\begin{tabular}[c]{@{}c@{}}\rotatebox{90}{\textbf{New Schemes}}\end{tabular}}
& Multiple Component Feedback {\color{black}[48][49][51]}& Feedback overhead can be reduced. & \makecell[l]{May need additional design (e.g., feedback \\sizes or intervals indications).} \\
\cline{2-4}
& Finite Feedback {\color{black}[52]-[55]}& Feedback overhead can be reduced. & \makecell[l]{May need additional design (e.g.,well-designed\\ codebook).}\\
\cline{2-4}
& Two-way Channel Sounding {\color{black}[56]}& Feedback overhead can be reduced. & \makecell[l]{May need additional design (e.g., especial \\sounding signal design).}\\
\cline{2-4}
& Enhanced Implicit Feedback {\color{black}[49][63]}& \makecell[l]{Improving from IEEE 802.11n and with \\low network overhead and latency.} & Need calibration.\\ \hline
\end{tabular}
\end{table*}
\section{Multiple Input Multiple Output(MIMO) Enhancement}
The use of 16 spatial streams has been discussed as an attractive MIMO feature of EHT. The increasing number of spatial streams from the current eight in IEEE 802.11ax to sixteen could theoretically double the transmission data rate. However, this comes with an attendant increase in the amount of sounding and feedback needed. Straightway reusing the existing sounding and feedback mechanisms defined in IEEE 802.11ax is not adequate to support 16 spatial streams. Therefore, this section mainly emphasizes the enhanced feedback reduction schemes to support 16 spatial streams as well as some new schemes after a brief introduction of existing schemes.

\subsection{Current Channel Sounding and Feedback Reduction Schemes}

The channel sounding mechanism is crucial to acquire accurate CSI for precoding the transmit signal in MIMO transmissions. There are two typical methods to acquire CSI: the implicit sounding supported in \cite{ref44} which relies on channel reciprocity to estimate the CSI at the receiver by using the CSI at the transmitter, and the explicit sounding supported in IEEE 802.11 specifications \cite{ref3}, \cite{ref7}, \cite{ref45,ref46,ref47} which requires the receiver to make CSI estimation and timely feedback the CSI to the transmitter.

Reducing the CSI feedback overhead is a key issue to improve the channel sounding efficiency, as lengthy feedback delays interfere with the timeliness of sounding. Implicit feedback overhead can be eliminated in reciprocal systems. On the premise of no performance loss, explicit feedback overhead can be reduced by limiting the amount of feedback information. In general, there are four kinds of explicit feedback reduction methodologies in current IEEE 802.11 systems: \emph{$\phi$} Only Feedback-feed back \emph{$\phi$} only in N $\times$ 1 (N is the number of transmitting antennas) transmission and assume a fixed \emph{$\psi$} (IEEE 802.11ah \cite{ref46}), Time Domain Channel Feedback (IEEE 802.11ad \cite{ref7} / IEEE 802.11ay \cite{ref47}), Given's Rotation-feed back time or frequency in Given's Rotation angles \cite{ref7}, and Angle Quantization-Given's angle (\emph{$\phi$},\emph{$\psi$}) is quantized evenly \cite{ref7}, where (\emph{$\phi$},\emph{$\psi$}) are the angles of the premultiplication matrices and Given's rotation matrices used in compressing the right singular matrix of the channel for feedback. Besides, IEEE 802.11ax \cite{ref3} recommends three other feedback reduction ways: increasing the tone grouping size during feedback, allowing partial bandwidth feedback over a range of RUs, and allowing feedback of the SNR of an RU.
\subsection{Enhanced Feedback Reduction Schemes}
Over the years, more spatial streams and better spatial multiplexing capabilities have been consistently expected for APs. However, as the total number of spatial streams increases up to 16, a large amount of sounding and feedback information may hinder gains of MIMO transmissions. Moreover, in the case of multi-AP scenarios in Section V, more feedback information is required since the STA may need to send feedback to each AP. Based on the above reasons, the most useful feedback overhead reduction methods for 16 spatial streams are required to be investigated. Table V summarizes many kinds of enhanced and new schemes for reducing the feedback overhead of channel acquisition. These schemes have their advantages and disadvantages, but which schemes are proper sounding protocols for 16 spatial streams need further evaluations.

\subsubsection{Enhanced explicit feedback}
EHT may improve the current explicit feedback reduction methodologies in IEEE 802.11 using any one of {\emph{$\phi$} Only Feedback, Time Domain Channel Feedback, Differential Given's Rotation, and Variable Angle Quantization. Meanwhile, we also highlight new explicit schemes, which may need additional designs, such as feedback schemes indication and channel sounding process designs. There are three kinds of new explicit feedback schemes to reduce feedback overhead, namely Multiple Component Feedback, Finite Feedback, and Two-way Channel Sounding.

\paragraph{\emph{$\phi$} Only Feedback \cite{ref48 ,ref49}}
In this method, we may keep the overhead the same and increase \emph{$b_\phi$}, and may also reduce the overhead by keeping \emph{$b_\phi$} the same and changing\emph{$b_\psi$}, where \emph{$b_\phi$} and \emph{$b_\psi$} are the number of quantized bits for Given's angle (\emph{$\phi$},\emph{$\psi$}). But this method can only work with a single data stream.

\paragraph{Time Domain Channel Feedback \cite{ref47}}
Feedback overhead can be saved when the number of significant taps may be much less than the number of tones. However, we may need to feed back the actual channel or the singular value decomposition components of the channel to enable the transformation of the channel to the frequency domain. It may require additional signaling to indicate extra matrix and taps position as they are not fixed as in frequency domain feedback. Therefore, there is a trade-off between increasing the feedback per tap and the smaller number of taps fed back in reducing the feedback overhead.

\paragraph{Differential Given's Rotation \cite{ref47,ref50}}
This differential feedback scheme significantly reduces feedback overhead by allowing each user to only feed back the difference in time or frequency in Given's Rotation angles relative to earlier feedback. For example, by using subtraction in the frequency domain, we can only send differential information between Given's rotation angles of baseline channel and next channel with a frequency separation of 4 sub-carriers. And this method requires additional processing and may also exist error propagation compared with the actual facts.

\paragraph{Variable Angle Quantization \cite{ref48,ref49}}
In the conventional method, Given's angle is quantized evenly. Depending on the channel state, we can use different quantization levels for different Given's rotation angles (\emph{$\phi_i$},\emph{$\psi_i$}) for explicit feedback reduction. And, it requires additional processing. Angle \emph{$\psi$} may vary over the distribution. To quantize the angles after Given's rotation, we may use different ranges for different angles or groups of angles: for each angle or groups of angles, the range $\Omega_\psi=[a,b]\subset{[0, \pi/2]}$.

\paragraph{Multiple Component Feedback \cite{ref48,ref49,ref51}}
This enhanced explicit method is to provide multi-component feedback by splitting feedback into multiple components. For example, one component has a larger size and is fed back at longer intervals, and another component is smaller and is sent back at shorter intervals, resulting in reducing overhead by combining feedback. It was an example of a scheme enabling a feedback granularity of fewer than 20 MHz in \cite{ref45}, and simulation result showed total feedback per transmission per user, i.e., the amount of feedback needed from a user to enable a successful transmission: can save approximate 90\% overhead per user per.
\begin{figure}[!htbp]
  \begin{center}
    \setlength{\abovecaptionskip}{-0.5cm}
    \scalebox{1.2}[1.2]{\includegraphics{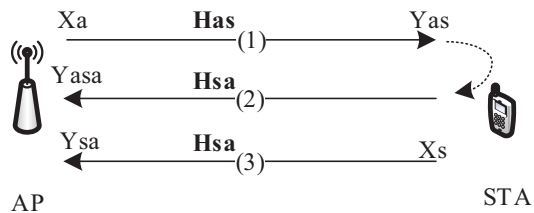}}
    \renewcommand{\figurename}{Fig.}
    {\color{black}\caption{Two-way channel sounding method for feedback reduction. The training signal Xa and Xs are known on both the AP side and the STA side. First, AP sends the training signal Xa via the channel Has to the STA. After receiving this as Yas, the STA returns Yas to the AP. The AP receives Yasa. Second, the AP receives Xs via the channel Hsa, as Ysa. From the two-way transmission, the AP estimates the channel impulse response Hsa and then the channel impulse response Has.}}
  \end{center}
\end{figure}
\begin{figure}[!t]
  \begin{center}
    \setlength{\abovecaptionskip}{-0.5cm}
    \scalebox{1.3}[1.3]{\includegraphics{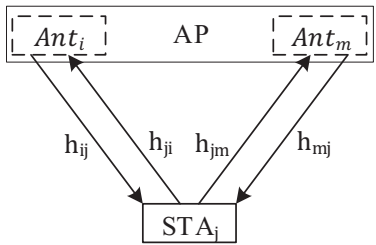}}
    \renewcommand{\figurename}{Fig.}
    {\color{black}\caption{A model of the local AP calibration. The $Ant_i$, $Ant_m$ are antenna elements of the same AP, and $h_{ij}$ ($h_{mj}$) is the channel between antenna $i$ ($m$) at the AP side and antenna $j$ at the STA side.}}
  \end{center}
\end{figure}

\paragraph{Finite Feedback \cite{ref52,ref53,ref54,ref55}}
In wireless communication network, Finite Feedback is a proven technology can achieve near-optimal channel adaptation, which allows the receiver to send a small number of information bits about the channel conditions to the transmitter. For example, codebook widely used in cellular networks may be an effective solution for feedback reduction. A finite feedback system may feed back codeword from a well-designed codebook, and the overall feedback may be reduced based on the size of the codebook \cite{ref52}. Besides, the high-precision compressive feedback technology by applying sparse approximation and compression may reduce overhead and resource consumption \cite{ref53,ref54,ref55}, which quantifies the channel vector by the linear combination of several unit vectors, then , then near-real CSI can be obtained, since the linear combination can be perfectly recovered by compressed sensing. This feedback needs to exploit signal sparsity characteristics in signal processing.

\paragraph{Two-way Channel Sounding \cite{ref56}}
Withers \emph{et al.} \cite{ref56} proposed a two-way channel sounding scheme shown in Fig. 11, that is, the AP sends training signals (Xa) to the STA via the channel (Has), and the STA repeatedly sends the received signals (Yas) back to the AP via the channel (Has). From this round-trip training signal (Yasa), together with the one-way training signal (Ysa) from the STA, the AP enables estimating its outgoing channel. This method has low complexity at the STA.

\begin{figure*}[!t]
\centering
    \setlength{\abovecaptionskip}{-0.3cm}
    \scalebox{0.85}[0.85]{\includegraphics{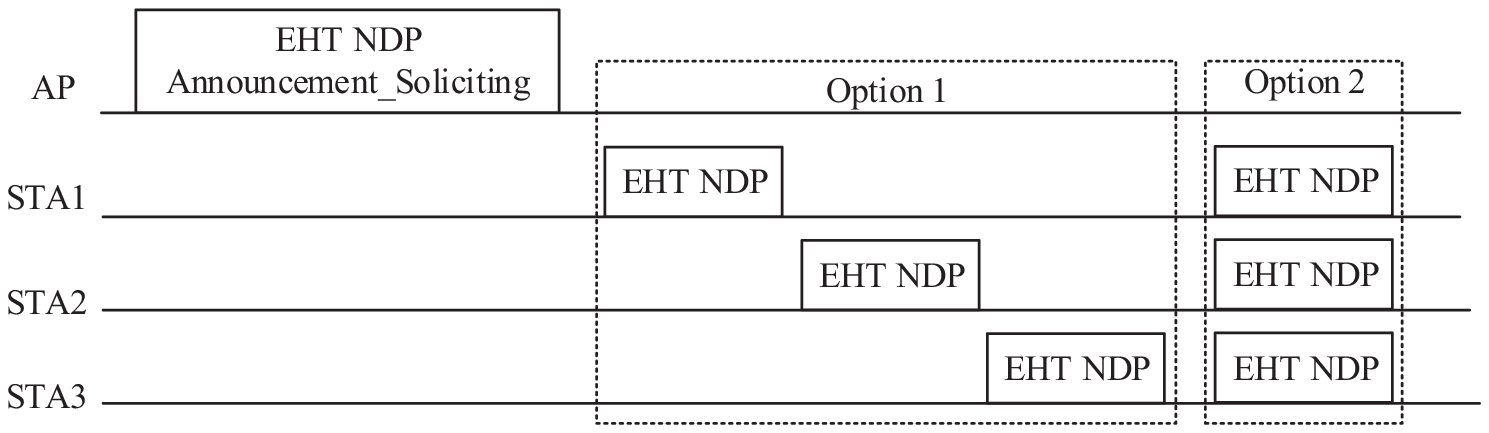}}
    \renewcommand{\figurename}{Fig.}
    \caption{Enhanced channel sounding with implicit feedback. After receiving the NDP Announcement frame, STAs will perform the sequential NDP transmission process or simultaneous NDP transmission process in OFDMA/UL MU-MIMO way.}
\end{figure*}
\subsubsection{Enhanced implicit feedback}
Implicit feedback can avoid overhead of the CSI feedback relying on the fact that UL and DL channels have identical impulse response in the same coherence interval. However, radios typically have slowly (and randomly) varying, non-reciprocal, impairments in the baseband-to-radio-frequency (RF) and RF-to-baseband chains. As a result, the actual DL baseband channel is not equal to the actual UL baseband channel unless this mismatch is explicitly compensated through calibration. In addition to the complex interactive methods specified in IEEE 802.11n \cite{ref44}, calibration can be solved today based on tremendous research efforts \cite{ref57 ,ref58,ref59}, including those based on smoothing and predistortion compensation. New developed local AP calibration may be applied where the STA is not required to be involved in the calibration process \cite{ref60,ref61,ref62}, which is no need for exchanging reference signals and channel information with other devices. This works based on the fact that in beamforming/linear precoding, it is sufficient for antennas to have a relatively accurate channel estimation \cite{ref62}. In Fig. 12, $Ant_i$, $Ant_m$ are antenna elements of the same AP, and $h_{ij}$ ($h_{mj}$) is the channel between antenna $i$ ($m$) at the AP side and antenna $j$ at the STA side. The calibration factors ($K$) for antennas $i$, $m$ are
\[{K_{i - j}} = \frac{{{h_{ij}}}}{{{h_{ji}}}}=\frac{{{{{t_i}} \mathord{\left/
 {\vphantom {{{t_i}} {{r_i}}}} \right.
 \kern-\nulldelimiterspace} {{r_i}}}}}{{{{{t_j}} \mathord{\left/
 {\vphantom {{{t_j}} {{r_j}}}} \right.
 \kern-\nulldelimiterspace} {{r_j}}}}}\eqno(1)\]
\[{K_{m - j}} = \frac{{{h_{mj}}}}{{{h_{jm}}}}=\frac{{{{{t_m}} \mathord{\left/
 {\vphantom {{{t_m}} {{r_m}}}} \right.
 \kern-\nulldelimiterspace} {{r_m}}}}}{{{{{t_j}} \mathord{\left/
 {\vphantom {{{t_j}} {{r_j}}}} \right.
 \kern-\nulldelimiterspace} {{r_j}}}}}\eqno(2)\]
\\By dividing equation (1) by equation (2),we can get the ratio of the calibration factors ($K$) for antennas $i$, $m$\\
\[\frac{{{K_{m - j}}}}{{{K_{i - j}}}} = \frac{{{{{t_m}} \mathord{\left/
 {\vphantom {{{t_m}} {{r_m}}}} \right.
 \kern-\nulldelimiterspace} {{r_m}}}}}{{{{{t_i}} \mathord{\left/
 {\vphantom {{{t_i}} {{r_i}}}} \right.
 \kern-\nulldelimiterspace} {{r_i}}}}}\eqno(3)\]
if\[{\rm{ }}{K_{i - j}} = {c_1} = 1\eqno(4)\]
\\
then\[{c_m} = {K_{m - j}} = \frac{{{{{t_m}} \mathord{\left/
{\vphantom {{{t_m}} {{r_m}}}} \right.
\kern-\nulldelimiterspace} {{r_m}}}}}{{{{{t_i}} \mathord{\left/
{\vphantom {{{t_i}} {{r_i}}}} \right.
\kern-\nulldelimiterspace} {{r_i}}}}} = \frac{{{h_{mi}}}}{{{h_{im}}}}\eqno(5)\]
\\
where $t_i$ and $r_i$ are transmitter and receiver at AP/ STA$_j$, $c_m$ is the relative calibration factor for antenna element $m$, $h_{im}$ is propagation channel between antenna $i$ and $m$ at the AP side, calibration factors ($K$) are applied on channel matrix and then beamforming vector is calculated.

Doostnejad \emph{et al.} \cite{ref62} evaluated Least Squares \cite{ref60} and Argos \cite{ref61} schemes in the local AP calibration, where Least Squares calibration at AP provides improvement in calibration accuracy and may result in less than 3 deg residual error at each element. Also, the impact of calibration error on MU-MIMO beamforming is evaluated. For a higher number of users in DL MU-MIMO beamforming, AP calibration error has to be maintained in lower range (less than 3 deg).\par
Based on those calibration methods mentioned above, it is assumed that the calibration between the receiver and the sender has been implemented, and an enhanced implicit channel scheme is discussed in \cite{ref49,ref63}. The detailed process of channel sounding is shown in Fig. 13. For MU-MIMO scenario, there is a trade-off between the overhead of implicit method (option 1) and explicit method, which mainly depends on the number of STAs, the feedback duration, and the duration of UL sounding frames. For example, as the number of STAs increases, the number of frames transmitted in UL also increases. Further, it can be observed that option 2 can effectively reduce the CSI feedback overhead of more than one STA compared with option 1.

\subsection{Summary of MIMO Enhancements}
This section mainly emphasizes the enhanced MIMO protocol to support a maximum of 16 spatial streams as well as some new channel sounding schemes after a brief introduction of existing channel sounding schemes. Both enhanced explicit and implicit channel sounding schemes are surveyed in this section. The enhanced explicit channel sounding schemes generally require additional designs (e.g., codebook design, quantization/compression processing) to reduce sounding overhead, but may be used as the mandatory mechanism for backward compatibility. The enhanced implicit channel sounding relying on channel reciprocity can offer significant potential for reducing sounding overhead. However, large calibration error in implicit channel sounding leads to non-effective MIMO transmissions. In addition to designing effective algorithms to smooth or compensate calibration error, it is critical to improve the hardware symmetry between the transmitter and the receiver.
\begin{table*}[!htbp]
\arrayrulecolor{black}
\renewcommand\arraystretch{1.3}
\centering
{\color{black}\caption{Summary of Existing Studies on Multi-AP Transmission and Performance Analysis.}}
\begin{tabular}{|c|l|}
\hline
\multicolumn{1}{|l|}{{\color{black} \textbf{}}}                                                                & \multicolumn{1}{c|}{{\color{black} \textbf{Contributions}}}                                                                                                                                                                                                                         \\ \hline
{\color{black} }                                                                                               & {\color{black} \begin{tabular}[c]{@{}l@{}}\-- Categorization of Multi-AP coordination including CBF, dynamic AP selection, and JTX, Multi-AP \\~{} sounding and joint transmission procedure {[}66{]}\end{tabular}} \\ \cline{2-2}
{\color{black}}                                                                                               & {\color{black} \begin{tabular}[c]{@{}l@{}}\-- Distributed MU-MIMO architecture design for Multi-AP, distributed channel access function, \\~{} Multiple MAC sublayers enhancements for Multi-AP {[}67{]}\end{tabular}} \\ \cline{2-2}
{\color{black}}                                                                                               & {\color{black}\begin{tabular}[c]{@{}l@{}}\-- Multi-AP phase synchronization, CSI measurement, edge user identification {[}68{]}\end{tabular}}                                                                             \\ \cline{2-2}
{\color{black}}                                                                                               & {\color{black}\begin{tabular}[c]{@{}l@{}}\-- Geographical location-based edge user identification {[}69{]}{[}70{]}\end{tabular}}
\\ \cline{2-2}
{\color{black}}                                                                                               & {\color{black}\begin{tabular}[c]{@{}l@{}}\-- SINR-based edge user identification {[}71{]}\end{tabular}}
\\ \cline{2-2}
{\color{black}}                                                                                               & {\color{black}\begin{tabular}[c]{@{}l@{}}\-- A reference channel as a reference clock for phase synchronization {[}73{]}{[}74{]}{[}75{]}{[}76{]}\end{tabular}}
\\ \cline{2-2}
{\color{black}}                                                                                               & {\color{black}\begin{tabular}[c]{@{}l@{}}\-- Procedure with the Multi-AP sounding, Multi-AP selection, and Multi-AP transmission {[}77{]}\end{tabular}}
\\ \cline{2-2}
{\color{black}}                                                                                               & {\color{black}\begin{tabular}[c]{@{}l@{}}\-- Multi-AP explicit sounding protocol for JTX and CBF {[}78{]}\end{tabular}}
\\ \cline{2-2}
{\color{black}}                                                                                               & {\color{black}\begin{tabular}[c]{@{}l@{}}\-- Explicit sounding procedure for Multi-AP coordination {[}79{]}\end{tabular}}
\\ \cline{2-2}
{\color{black}}                                                                                               & {\color{black}\begin{tabular}[c]{@{}l@{}}\-- Collaborative cluster selection for Multi-AP transmission {[}80{]}{[}81{]}\end{tabular}}
\\ \cline{2-2}
{\color{black}}                                                                                               & {\color{black}\begin{tabular}[c]{@{}l@{}}\-- CSR combined with C-OFDMA {[}83{]}\end{tabular}}                                                  \\ \cline{2-2}
\multirow{-12}{*}{{\color{black} \textbf{\begin{tabular}[c]{@{}c@{}}Multi-AP Transmission\\ Designs\end{tabular}}}} & {\color{black}\-- Coordinated association/handover, coordinated timing scheduling {[}88{]}}                                                                                                                                                                                                \\ \hline
{\color{black}}                                                                                               & {\color{black}\-- Feasibility analysis of joint beamforming with different non-idealities/impairments {[}72{]}}                                                                                                                                                            \\ \cline{2-2}
{\color{black}}                                                                                               & {\color{black} \begin{tabular}[c]{@{}l@{}}\-- Simplified throughput gain analysis of C-OFDMA and CBF {[}82{]}\end{tabular}}                                                                                                     \\ \cline{2-2}
{\color{black} }                                                                                               & {\color{black}\begin{tabular}[c]{@{}l@{}}\-- Simplified simulation on throughput gain obtained by using CSR in the UL and DL {[}83{]}\end{tabular}}                                                                                        \\ \cline{2-2}
{\color{black} }                                                                                               & {\color{black}\begin{tabular}[c]{@{}l@{}}\-- Controlling transmission power for CSR and simulation analysis on throughput gain of CSR {[}85{]}\end{tabular}} \\ \cline{2-2}
{\color{black}}                                                                                               & {\color{black}\begin{tabular}[c]{@{}l@{}}\-- Performance analysis on sum throughput gain of CBF, single AP and CSR {[}87{]}\end{tabular}}
\\ \cline{2-2}
{\color{black}}                                                                                               & {\color{black}\begin{tabular}[c]{@{}l@{}}\-- Evaluation of coordinated Multi-AP in uplink whether the scheme is attractive and feasible or not {[}89{]}\end{tabular}}
\\ \cline{2-2}
\multirow{-6}{*}{{\color{black} \textbf{\begin{tabular}[c]{@{}c@{}}Performance Analysis\end{tabular}}}}     & {\color{black}\begin{tabular}[c]{@{}l@{}}\-- Performance investigation on CBF, JTX and Non-coordinated transmission {[}92{]}\end{tabular}}                                                 \\ \hline
\end{tabular}
\end{table*}

\section{Multi-AP Coordination}
With the ever-growing of mobile users and thereby increasing demand, co-channel interference becomes unbearable in dense wireless networks. Collaboration between adjacent APs, such as sharing necessary scheduling information and CSI, is a promising approach to improve the utilization of limited radio resources. In this section, we present the multi-AP network and a general multi-AP transmission procedure, emphasize several modes of multi-AP transmission including C-OFDMA, CSR, CBF and JXT, and summarize existing studies on multi-AP coordination in Table VI.

\subsection{Multi-AP Network Architecture}
Under typical multi-AP network scenarios, such as home network, enterprise network and commercial network, an AP has to communicate with its neighboring APs for coordination, causing substantial signaling overhead and processing complexity. In this regard, centralized network architectures, such as cloud architecture \cite{ref64} and software-defined network (SDN) \cite{ref65}, have considerable potential to reduce the complexity of synchronization and coordination process among physically independent APs. As shown in Fig. 14, the multi-AP system has a master AP (M-AP) and multiple slave APs (S-APs), where the M-AP as the coordinator of all APs is helpful in multiple APs' management and resource scheduling, and the S-APs participate in the multi-AP transmissions \cite{ref66}.

\setcounter{figure}{13}
\begin{figure}[t]
  \begin{center}
    \setlength{\abovecaptionskip}{-0.2cm}
    \scalebox{1.3}[1.3]{\includegraphics{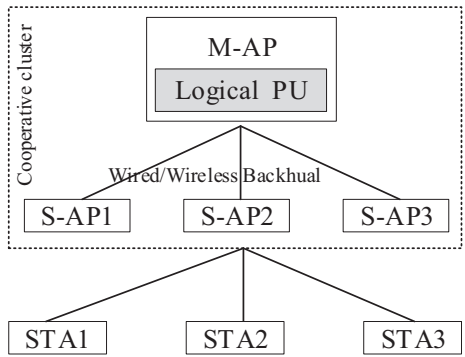}}
    \renewcommand{\figurename}{Fig.}
    \caption{An example of the multi-AP architecture, where a group of APs are connected by wired or wireless links and the logical processing unit (PU) located at the M-AP is used to coordinate/control the related multi-AP operations of multiple distributed APs.}
  \end{center}
\end{figure}\par

Besides, high-capacity, low-latency wired (e.g., fibre) or wireless (e.g., millimeter-wave) backhaul links are also needed to exchange coordination-related information and service data in real-time among multiple APs \cite{ref66}\cite{ref67}. To realize efficient process for multi-AP transmissions, a logical processing unit (PU) \cite{ref67} could be added to the multi-AP network to coordinate the related multi-AP operations of multiple distributed APs , such as managing resources of all APs, managing the Carrier Sense Multiple Access/Collision Avoidance (CSMA/CA) function of all APs, and coordinating the transmission of all APs, etc. In general, if the STA is far away from the interfering APs, it will not suffer from relatively significant interference, and its communication quality can be guaranteed without multi-AP collaboration. Consequently, specific criteria are needed to identify which STAs are edge users. Random algorithm \cite{ref68} and geographical location \cite{ref69}\cite{ref70} were used for selecting users. However, APs are normally randomly deployed without any planning in WLAN. It is not reasonable to distinguish users by random algorithm or geographical location. Basically, the center user differs from the edge user according to the Signal to Interference plus Noise Ratio (SINR).  Thus, the AP can determine the STA to be an edge STA if the SINR calculated is less than the pre-set threshold, or to be a center user otherwise \cite{ref71}.\par
\setcounter{figure}{14}
\begin{figure*}[t]
  \begin{center}
    \setlength{\abovecaptionskip}{-0.5cm}
    \scalebox{0.7}[0.7]{\includegraphics{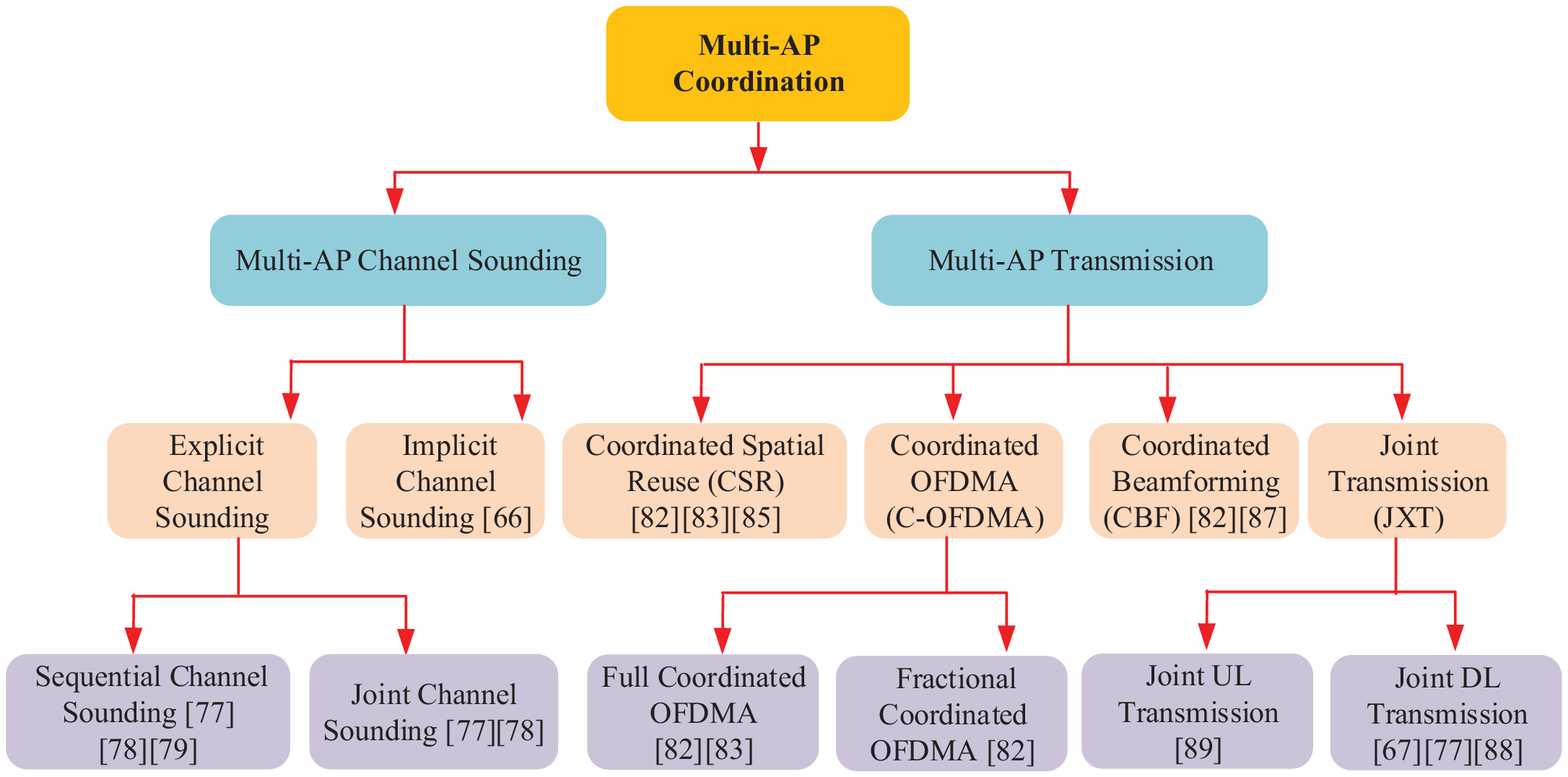}}
    \renewcommand{\figurename}{Fig.}
    {\color{black}\caption{Overview of the multi-AP coordination.}}
  \end{center}
\end{figure*}
Unlike the traditional AP where all transmitter antennas share the same oscillator, multiple distributed APs have their local oscillators with independent carrier frequency offsets (CFOs). How to synchronize the oscillators of multiple distributed APs is a big challenge for the multi-AP network. Imperfect synchronization has some residual CFOs remain. Even tiny residual CFOs, with the phases of drift away after tens of symbols, will result in decoding errors. Simulations showed in \cite{ref72} that phase buildup from even a low 20 Hz residual CFO across APs causes peak throughput degradation significantly. To offset the impact of residual CFO, the AP/STA can use mid-ambles inserted in a PPDU transmission to update/replace channel estimation with accumulated phase in fast varying channels, i.e., channels with high Doppler shift. As what is done in IEEE 802.11ax, the M-AP also can leverage a trigger frame to enable S-APs to initially sync and subsequently re-sync their timing, CFO and phase in each part of the process, including sounding and every multi-AP transmission thereafter. It is easy to do based on the legacy preamble. In addition, using a reference channel as a reference clock for the phase synchronization purpose appears in AirSync \cite{ref73}, Vidyut \cite{ref74}, Poster \cite{ref75} and AirShare \cite{ref76}. Based on the wireless channel or the powerline backbone, however, the reference channel requires extra hardware complexities on each AP. Without additional hardware complexities, a collaborative tracking scheme was proposed in \cite{ref68} to track phase drifts at the symbol level. The core idea is to have one reference AP to monitor ongoing transmissions and compute the phase drifts of each data symbol and feedback these estimations to all APs via an Ethernet connection. Based on this feedback, APs can dynamically adjust their signal phases to ensure strict alignment.
\setcounter{figure}{15}
\begin{figure*}[t]
  \begin{center}
    \setlength{\abovecaptionskip}{-0.5cm}
    \scalebox{0.9}[0.9]{\includegraphics{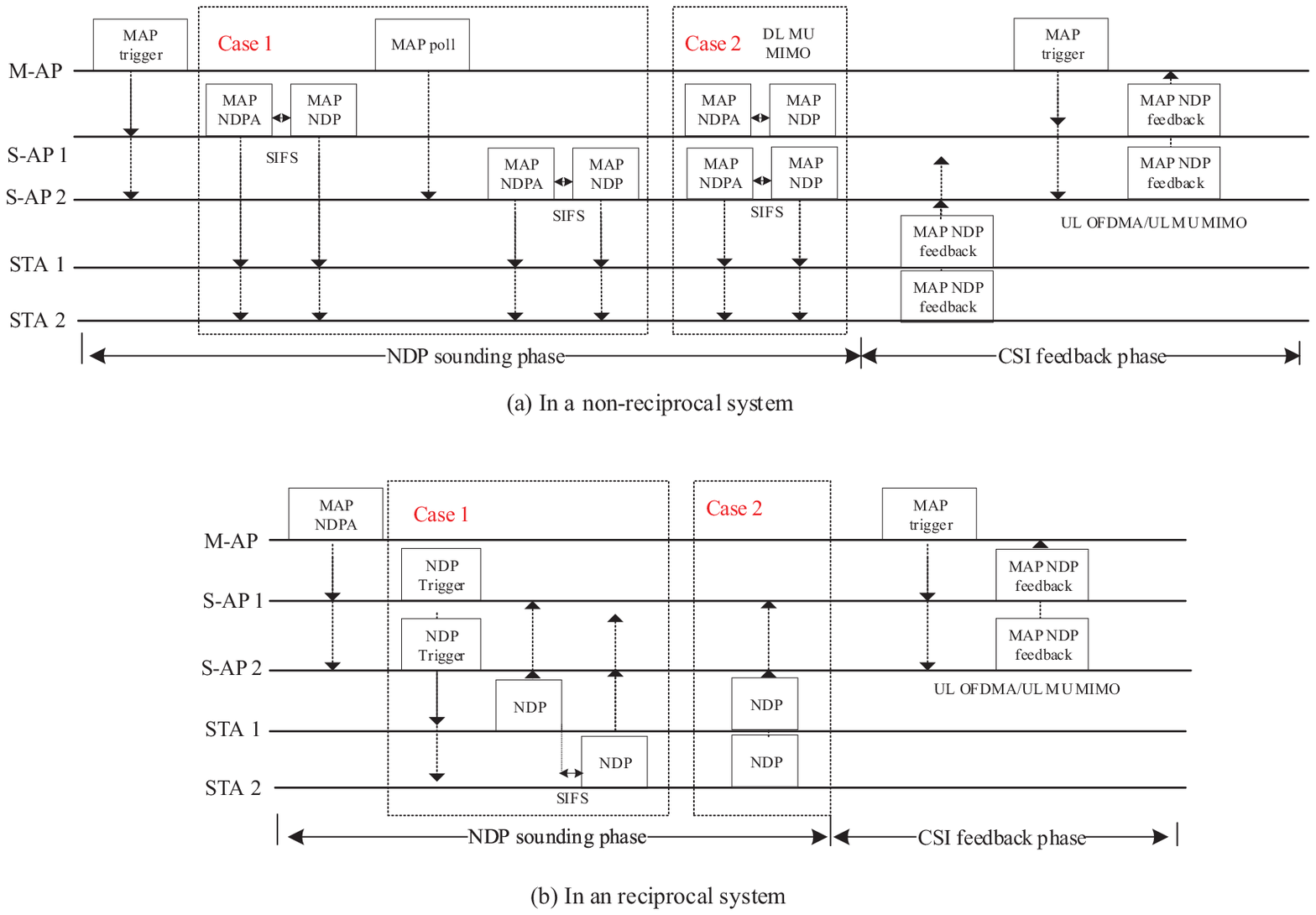}}
    \renewcommand{\figurename}{Fig.}
    {\color{black}\caption{Acquisition of CSI between the multiple APs and STAs in the non-reciprocal/reciprocal system. (a) In the non-reciprocal system, the coordinate AP (M-AP) first initiates the multi-AP sounding by sending the NDPA frame, then multiple S-APs will send NDP simultaneously/sequentially following the sequence specified in the NDPA frame, and finally the STAs feedback CSI in the OFDMA/UL MU-MIMO way. (b) In the reciprocal system, multiple STAs transmit NDP sounding frame simultaneously/sequentially in response to the soliciting NDP Trigger frame containing information indicating the timing at which the STAs respectively to transmit NDP sounding frames.}}
  \end{center}
\end{figure*}

\begin{table*}[!t]
\arrayrulecolor{black}
\centering
\renewcommand\arraystretch{1.2}
\setlength{\abovecaptionskip}{0.cm}
\setlength{\belowcaptionskip}{-0.cm}
{\color{black}\caption{Analysis and comparison of four multi-AP transmission modes.}}
\begin{tabular}{|c|c|c|c|c|}
\hline
{\color{black}\textbf{}}                                                                   & {\color{black}\textbf{C-OFDMA } }                                                               & {\color{black}\textbf{CSR}  }                                                                                          & {\color{black}\textbf{CBF} }                                                                                           & {\color{black}\textbf{JTX} }                                                                                  \\ \hline
\textbf{\begin{tabular}[c]{@{}c@{}}{\color{black}Application Scenario} \\ {\color{black}}\end{tabular}}             & {\color{black}DL and/or UL }                                                           & {\color{black}DL and/or UL }                                                                                  & {\color{black}DL}                                                                                             & {\color{black}DL and/or UL}                                                                          \\ \hline
\textbf{\begin{tabular}[c]{@{}c@{}}{\color{black}Coordinated} \\ {\color{black}Domain}\end{tabular}}               & {\color{black}Time/Frequency }                                                         & {\color{black}Power/Spatial}                                                                                    & {\color{black}Spatial  }                                                                                        & {\color{black}Spatial }                                                                                \\ \hline
\textbf{\begin{tabular}[c]{@{}c@{}}{\color{black}Sharing Info} \\ {\color{black}between APs}\end{tabular}}         & {\color{black}CSI, Time/Frequency}                                                     & \begin{tabular}[c]{@{}c@{}}{\color{black}Spatial Reuse Info,} \\ {\color{black}Traffic Info}\end{tabular}                 & {\color{black}CSI}                                                                                            & {\color{black}CSI, User Data}                                                                        \\ \hline
\textbf{\begin{tabular}[c]{@{}c@{}}{\color{black}Number of Serving-AP} \\ {\color{black}for One STA}\end{tabular}} & {\color{black}Single AP }                                                                 & {\color{black}Single AP }                                                                                        & {\color{black}Single AP  }                                                                                       & {\color{black}Multiple APs  }                                                                        \\ \hline
\textbf{\begin{tabular}[c]{@{}c@{}}{\color{black}Synchronization or} \\{\color{black}Backhaul Requirements}
\\ {\color{black}between APs}\end{tabular}}      & \begin{tabular}[c]{@{}c@{}}{\color{black}Symbol Level} \\ {\color{black}Synchronization}\end{tabular} & \begin{tabular}[c]{@{}c@{}}{\color{black}PPDU Level} \\ {\color{black}Synchronization}\end{tabular}                          & \begin{tabular}[c]{@{}c@{}}{\color{black}Symbol Level} \\ {\color{black}Synchronization}\end{tabular}                        & \begin{tabular}[c]{@{}c@{}}{\color{black}Tight Time/Frequency/}\\ {\color{black}Phase Synchronization, }\\ {\color{black}Backhaul Requirements }\end{tabular}\\ \hline
\textbf{\begin{tabular}[c]{@{}c@{}}{\color{black}Coordination} \\ {\color{black}Complexity}\end{tabular}}          & {\color{black}Low/Medium }                                                             & {\color{black}Low}                                                                                            & {\color{black}Medium  }                                                                                       & {\color{black}Very High}                                                                             \\ \hline
{\color{black}\textbf{Benefit} }                                                                    & {\color{black}Interference Mitigation}                                                 & \begin{tabular}[c]{@{}c@{}}{\color{black}Improved Spatial Reuse,}\\      {\color{black}Interference Mitigation}\end{tabular} & \begin{tabular}[c]{@{}c@{}}{\color{black}Improved Spatial Reuse,}\\      {\color{black}Interference Mitigation}\end{tabular} & \begin{tabular}[c]{@{}c@{}}{\color{black}Improved Spatial Reuse,} \\ {\color{black}Higher Reliability}\end{tabular} \\ \hline
\end{tabular}
\end{table*}

\subsection{Multi-AP Transmission Procedures}
IEEE 802.11ax only supports transmission to/from a single AP and spatial sharing between APs and STAs, while EHT extends its ability to multi-AP transmissions initiated by AP or non-AP STA based on multi-AP scenarios and/or QoS requirements, such as when requiring higher efficiency, higher peak throughput, and lower latency and jitter. In a multi-AP network, to take full advantage of multi-AP transmissions, the multi-AP channel sounding is required to provide CSI from STAs to APs participating in a multi-AP transmission. As shown in Fig. 15, two typical channel sounding methods are considered for multi-AP transmissions, i.e., explicit channel sounding with serious concerns on computational complexity and CSI feedback overhead, and implicit channel sounding with calibration requirement of receive/transmit chains. With respect to the different coordination complexity, four types of multi-AP transmission schemes have been discussed in the EHT task group including C-OFDMA, CSR, CBF and JXT. In this part, we will present details of multi-AP sounding schemes and multi-AP transmission schemes.

\subsubsection{Multi-AP channel sounding procedure}
The multi-AP sounding procedure should be performed beforehand to acquire the CSI between multiple APs and STAs. In Fig. 16, the M-AP transmits the trigger/Null Data Packet Announcement (NDPA) frame to the S-APs to initiate the explicit or implicit sounding procedure. In principle, the NDP transmission should be defined considering time overhead, channel estimation accuracy and computational complexity both on the AP side and STA side. In a non-reciprocal multi-AP system \cite{ref77,ref78,ref79}, the explicit channel sounding process, in which STAs estimate channels and feed back CSI to the AP, can adopt sequential NDP transmissions or simultaneous NDP transmissions as shown in Fig. 16 (a). The polling-based NDP sequential transmission does not require synchronization between multiple S-APs, but the time overhead of the sounding procedure gradually increases as the total number of participating S-APs increases. The simultaneous transmission using tone selection/DL MU-MIMO based on the Long Training field has a stringent synchronization requirement among the S-APs. Even though the channel sounding protocol can reduce sounding overhead, it increases the CSI computation burden at the STA side, especially for the multi-AP scenario where a single AP can support up to 16 spatial streams. In a reciprocal multi-AP system, the implicit channel sounding scheme uses CSI at the AP side to estimate that at the STA, which requires calibration of receive/transmit chains and significantly reduces the CSI feedback overhead compared to explicit sounding. As shown in Fig. 16 (b), the STA should be solicited to send an MAP NDP frame to S-APs for implicit channel estimation \cite{ref66}.
After completing channel estimation, collecting CSI from all STAs/S-APs may be followed in the UL OFDMA/UL MU-MIMO way, which can reduce the time of collecting CSI compared to sequential CSI feedback. To further reduce high transmission time for collecting CSI, enhanced sounding schemes depicted in Section IV (e.g., Time Domain Channel Feedback and Multiple Component Feedback) can be applied to the CSI collection process from multi-AP and multi-STA. Based on the collected CSI, the M-AP decides which S-APs are best suitable for the multi-AP transmission and informs S-APs of being selected. If the selected S-AP cannot participate in the multi-AP transmission, and the M-AP reselects other potential S-AP(s). Additionally, there are many other potential strategies for multi-AP selection, e.g., static/semi-static/dynamic network-centric and user-centric collaborative cluster selection strategies \cite{ref80} and intelligent selection strategies based on deep reinforcement learning \cite{ref81}.
\begin{figure*}[!t]
  \begin{center}
    \setlength{\abovecaptionskip}{-0.2cm}
    \scalebox{0.46}[0.46]{\includegraphics{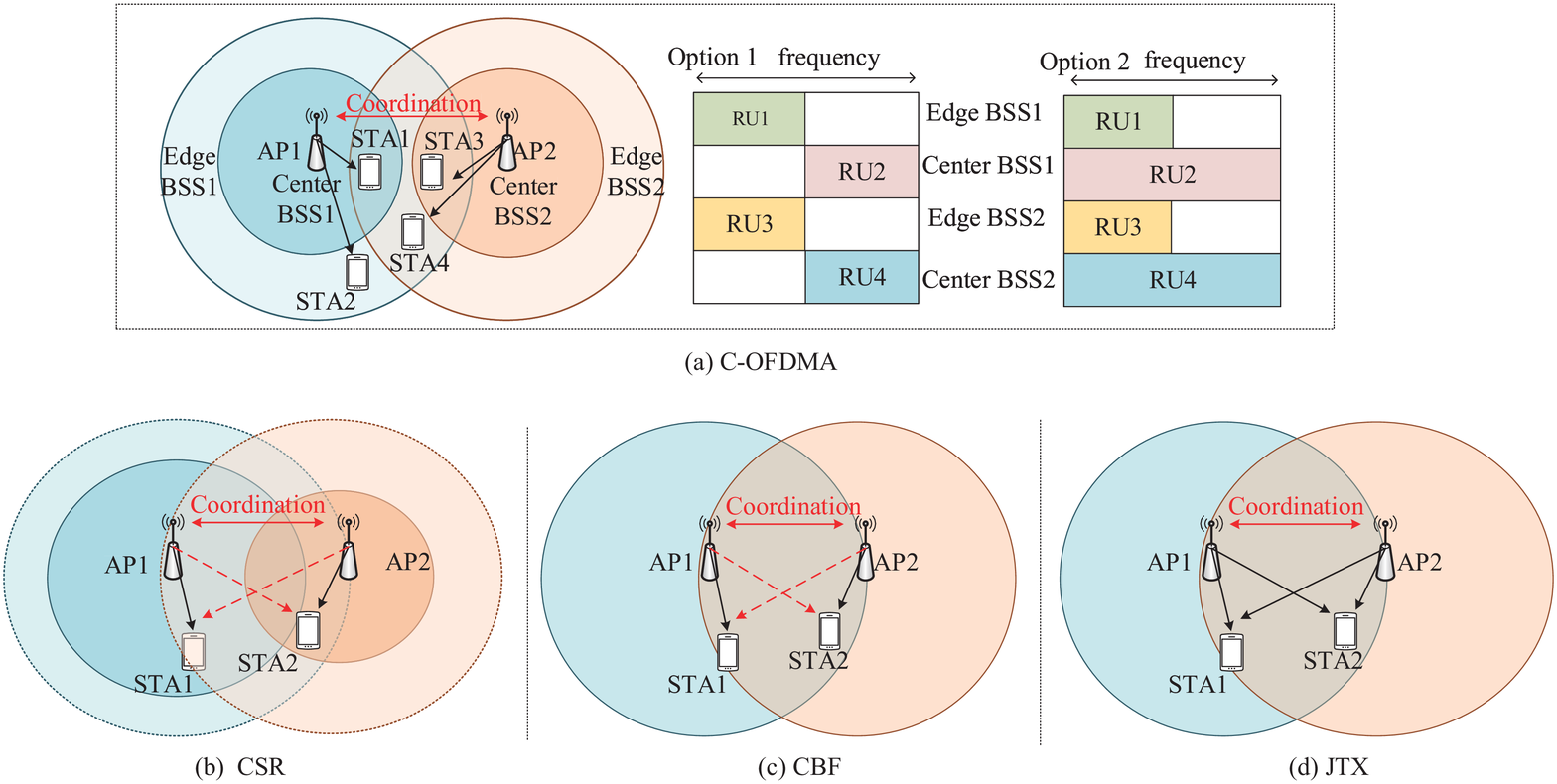}}
    \renewcommand{\figurename}{Fig.}
    {\color{black}\caption{An example of multi-AP transmission schemes. (a) In the C-OFDMA, each AP coordinates RUs for all its STAs or only interference limited (edge) STAs. (b) In the CSR, APs need to know the receiving STAs to decide the beamforming vector and transmit power. (c) In the CBF, each AP performs spatial domain nulling and limits its transmitted  interference to STAs in other BSSs while transmitting to its desired STA. (d) In the JTX, STAs can be jointly served by multiple distributed APs with the data for all participating STA.}}
  \end{center}
\end{figure*}

\subsubsection{Coordinated multi-AP transmission}
After performing the channel sounding procedure, the M-AP initiates the multi-AP transmission by sending a management frame to the final selected S-AP(s). There are many schemes of multi-AP transmission as summarized in the following Table VII, such as C-OFDMA, CSR, CBF, and JXT.\par

\paragraph{C-OFDMA}
OFDMA is a technology that divides the whole bandwidth into a series of OFDM subcarrier sets called RUs in IEEE 802.11ax and assigns different RUs to different users to achieve multiple accesses. EHT extends IEEE 802.11ax OFDMA from a single AP to multiple APs \cite{ref82}\cite{ref83}, which leads to efficient utilization of frequency resources across the network. For the Full Coordinated OFDMA illustrated in Fig. 17 (a), APs coordinate to share OFDMA resources for all STAs and enable different STAs to adopt mutually orthogonal time and frequency resources, thus avoiding RUs conflicts. This may result in-efficient allocation of resources as both APs are limited to the resource they are assigned to. To further improve the spectrum utilization, each AP should coordinate RUs for only interference limited (edge) STAs, i.e., Fractional Coordinated OFDMA. For example, non-interference limited STAs (center) may transmit/receive in all RUs, while interference limited STAs (edge) may transmit in coordinated RUs. In the C-OFDMA, one problem is that how to allocate appropriate RUs/channels to the recipient STAs. Two main resource allocation schemes based on frequency reuse (static) and cell-based coordination (dynamic) were reviewed in \cite{ref84}. The pre-allocation static frequency resources are generally easy to use but not easy to change resource allocation between adjacent APs according to the dynamic characteristics of the network. Dynamic allocation resources could fill the gaps in static resources, but it requires APs to timely know which channels/RUs are available.
\paragraph{CSR}
The objective of spatial reuse is to improve the system level performance, the utilization of medium resources and power saving in dense deployment scenarios through interference management. In the existing spatial reuse mechanism which is not a coordinated way, one AP can transmit data with the max transmission power, while the other APs should transmit data with the transmission power calculated by overlapping basic service set (OBSS) packet detect (PD) equation \cite{ref3}, which will result in some STAs getting too low SINR when some APs decrease transmission power and get into detecting OBSS signals. In addition, when multiple APs concurrently transmit data with the max transmission power, without coordination between APs, transmission power control might become ineffective to reduce interference. To further address the issues of spatial reuse in IEEE 802.11ax, EHT recommends to control the transmission power between APs in a coordinated way, such as periodically controlling power not detected at other APs as shown in Fig. 17(b) or controlling power in every transmission to maximize area throughput \cite{ref85}. Without considering the impact of the frame exchanges on the system overhead, controlling power in every transmission process will be the best way to improve the throughput when using spatial reuse. The simulation results in \cite{ref85} showed that CSR can achieve higher throughput gain compared with spatial reuse (OBSS\_PD) in IEEE 802.11ax. Under simulation settings in \cite{ref83}, through coordinated SR, more than 30\% throughput gain can be obtained in UL, and more than 20\% throughput gain can be obtained in DL. Furthermore, CSR can be combined with coordinated OFDMA \cite{ref82}\cite{ref83}. As shown in Fig. 17(a), each AP only coordinates RUs for the interference-limited edge STAs, while the center STAs far from the interfering AP may send/receive in all RUs without interference.

\paragraph{CBF}
In MIMO systems, beamforming is a wireless technology through which an AP can place spatial radiation nulls from and towards non-served STAs for interference suppression purposes. At present, beamforming is only performed by single AP independently in WLAN, resulting in uncontrollable inter-AP interference. To better spatial reuse and mitigate inter-cell interference, EHT recommends CBF \cite{ref82}\ each transmit signal on the intended STAs while doing the signal null at the non-served STAs, thus achieving multiple concurrent transmissions. To implement spatial radiation nulling in the multiple BSSs, one can use different ways as surveyed in \cite{ref86}. Simulation results in \cite{ref87} showed that CBF could provide major sum throughput gain over single AP and CSR. Moreover, CBF can be combined with coordinated OFDMA for the sake of more effective interference management. For example, STA 1, STA 2 and STA 3 are far from their own APs, STA 1 and STA 3 can share the same RU/channel through implementing CBF, and STA 1 and STA 2 can use different RUs/channels through implementing coordinated OFDMA.
\begin{figure}[!htbp]
  \begin{center}
    \setlength{\abovecaptionskip}{-0.5cm}
    \scalebox{0.95}[0.95]{\includegraphics{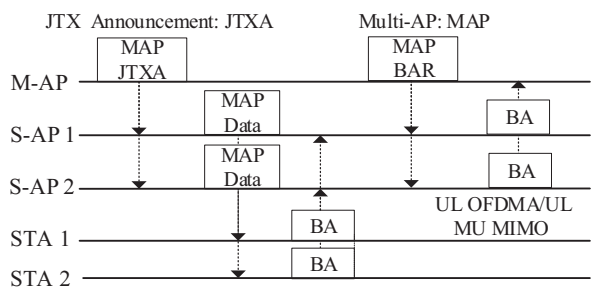}}
    \renewcommand{\figurename}{Fig.}
    {\color{black}\caption{An example of the DL JTX procedure. After receiving the JTXA frame from the M-AP, the S-APs know which data frames to send, what transmission parameters to use, and then the S-APs send data to the STAs. According to the newly designed BA/Ack transmission schemes, each STA transmits Ack/BA frame SIFS after data reception to its own associated S-AP or the M-AP after receiving the BA Request (BAR) frame.}}
  \end{center}
\end{figure}

\paragraph{JTX}
As shown in Fig. 17(d), JTX can be regarded as a virtual MIMO system consisting of multiple APs and multiple STAs \cite{ref67}\cite{ref77}\cite{ref88}\cite{ref89}. This technology also enables a fast association with an optimal AP and improves re-connection speed where users move around (e.g., office, hotspot, home scenario) \cite{ref88}. JTX targets at achieving joint transmissions/receptions between the non-collocated time/phase-synchronized APs and the time/phase-synchronized STAs. In Fig. 18, the M-AP sends the JTXA frame containing scheduling and other control information for joint transmission to the S-APs, and then all S-APs send data simultaneously to STAs. Each STA transmits Ack/BA frame after data reception to its own associated S-AP or to the M-AP after receiving the BA Request (BAR) frame. The UL JTX \cite{ref89} can provide higher reliability in various scenarios, and different approaches have been discussed in the EHT task group, including Distributed Interference Cancellation which improves UL data delivery, and Joint Reception which requires that all the APs jointly process the received data from all the STAs. For the DL JTX, one giant precoder for concurrent transmissions of multiple STAs can be applied over the combined array consisting of the transmitting antennas of all distributed APs. For a conventional AP, its co-located antennas share a common view of the channel status (idle/busy) relying on CSMA/CA. For JTX, the physical separation between APs leads to different views of the channel status. For this reason, there is a new centralized CSMA/CA mechanism in \cite{ref67}, in that each AP reports the CCA status to the PU, which deems the channel idle, and then a JTX is followed. When multiple APs continue to occupy the channels for JTX, the traditional APs cannot use channel resources because the channels are sensed as busy. Therefore, it is necessary to compromise the gain of centralized CSMA/CA and the fairness of traditional CSMA/CA. Furthermore, new architecture related topics are noteworthy (e.g., MAC/PHY splitting \cite{ref67}\cite{ref90}\cite{ref91}).

\subsection{Summary of the Multi-AP Coordination Operations}
This section describes issues related to the multi-AP coordination operations, including the multi-AP network, multi-AP channel sounding and multi-AP transmission. In a typical multi-AP scenario where a group of APs are connected by wired or wireless links, the centralized multi-AP network with a central node facilitates the inter-AP managements, such as resource scheduling, time synchronization and data sharing. However, the distributed multi-AP network without a central node has the main challenge of achieving tight synchronization among multiple APs, so that the multiple APs can work as one giant AP to perform the MIMO transmission or OFDMA operation. In the multi-AP network, to take full advantage of multi-AP coordination, explicit/implicit channel sounding is necessary to provide CSI from STAs to APs participating in the multi-AP transmission. However, there are serious concerns on the computational complexity and CSI feedback overhead in explicit sounding and calibration of receive/transmit chains in implicit sounding. To solve these problems, the multi-AP network can employ the enhanced sounding schemes depicted in Section IV (e.g., Two-way Channel Sounding, Time Domain Channel Feedback and Multiple Component Feedback, and Local AP Calibration).\par
As shown in Section V, the multi-AP transmissions fall into four categories: C-OFDMA, CSR, CBF and JXT. Under the simulation setting in \cite{ref92}, the throughput performance of CBF, JTX and non-coordinated transmission were compared and it showed different performance in the interference-limited region and noise-limited region. It means that the scheme of multi-AP transmission can be variable to meet the requirements of typical use scenarios. Therefore, we also need to explore how to choose an appropriate multi-AP transmission scheme based on what conditions to improve the system performance in more realistic and complex environments. Generally, in a long-term static environment where no new STAs and associated STAs access and associated users leave the network, respectively, the same multi-AP transmission scheme can be maintained for a long time without degrading the system performances. In dynamically changing environments where users frequently access or leave the network, we can adaptively switch the scheme among different multi-AP schemes according to the channel state, SINR or SNR. Furthermore, to mitigate interference and to enhance the system performances, we can consider combining the multi-AP scheme with other technologies, such as frequency/time allocation, spatial reuse, etc. In particular, machine learning, as an effective technology to reduce or even replace manual efforts in decision-making for wireless networks, can be leveraged to efficiently deal with the less-tractable problems, such as the selections of multi-AP modes and combined methods.
\begin{figure*}[!t]
  \begin{center}
    \setlength{\abovecaptionskip}{-0.5cm}
    \scalebox{1.15}[1.15]{\includegraphics{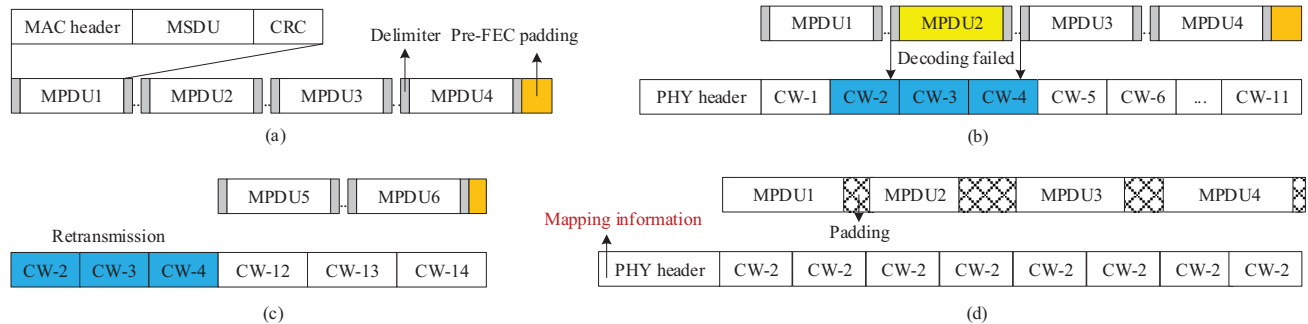}}
    \renewcommand{\figurename}{Fig.}
    \caption{Illustration of HARQ granularity. (a) HARQ at the A-MPDU level. (b) HARQ at the MPDU level. (c) HARQ at the CW level. (d) Padding for aligning the border of MPDU and CW.}
  \end{center}
\end{figure*}

\begin{table*}[!htbp]
\renewcommand\arraystretch{1.3}
\centering
{\color{black}\caption{Summary of Existing Studies on HARQ Granularity/Process and Performance Analysis.}}
\begin{tabular}{|c|l|}
\hline
\multicolumn{1}{|l|}{{\color{black} \textbf{}}}                                                                & \multicolumn{1}{c|}{{\color{black} \textbf{Contributions}}}                                                                                                                                                                                                                         \\ \hline
{\color{black} }                                                                                               & {\color{black} \begin{tabular}[c]{@{}l@{}}\-- Discussions on granularity at which HARQ can be supported and what changes would be necessary \\ ~{} at the PHY and MAC layer {[}93{]}\end{tabular}}                                                                                   \\ \cline{2-2}
{\color{black}}                                                                                               & {\color{black} \begin{tabular}[c]{@{}l@{}}\-- Alignment between LDPC codewords and MPDUs, effective solutions for supporting HARQ while\\ ~{} maintaining the existing LDPC and Block-ACK designs {[}94{]}\end{tabular}}                                                           \\ \cline{2-2}
{\color{black}}                                                                                               & {\color{black}\begin{tabular}[c]{@{}l@{}}\-- HARQ procedure from the perspective of PHY and MAC side {[}95{]}\end{tabular}}                                                                                      \\ \cline{2-2}
{\color{black}}                                                                                               & {\color{black}\begin{tabular}[c]{@{}l@{}}\-- Discussions about retransmission scheduling, HARQ control information and its exchange, HARQ\\ ~{} granularity for retransmission, HARQ ACK/NACK channel {[}97{]}\end{tabular}}                                                      \\ \cline{2-2}
\multirow{-5}{*}{{\color{black} \textbf{\begin{tabular}[c]{@{}c@{}}HARQ Granularity/\\ Process\end{tabular}}}} & {\color{black}\-- Alignment of CWs by applying padding method {[}98{]}{[}99{]}}                                                                                                                                                                                                \\ \hline
{\color{black}}                                                                                               & {\color{black}\--- Evaluation of the effect of decoding of the preamble on the goodput performance of HARQ {[}96{]}}                                                                                                                                                            \\ \cline{2-2}
{\color{black}}                                                                                               & {\color{black} \begin{tabular}[c]{@{}l@{}}\-- Evaluation on throughput gain of HARQ PCC and HARQ IR for LDCP encoding along with ARQ and \\~{} HARQ CC {[}100{]}\end{tabular}}                                                                                                     \\ \cline{2-2}
{\color{black} }                                                                                               & {\color{black}\begin{tabular}[c]{@{}l@{}}\-- HARQ complexity analysis of receiver LLR memory requirements, LDPC codeword processing \\ ~{} and MAC layer processing {[}101{]}\end{tabular}}                                                                                        \\ \cline{2-2}
{\color{black} }                                                                                               & {\color{black}\begin{tabular}[c]{@{}l@{}}\-- Simulation analysis of several issues regarding HARQ including, link adaptation method, effect of\\ ~{} preamble detection error, HARQ method (CC, IR), frequency diversity, and number of \\~{} retransmissions {[}102{]}\end{tabular}} \\ \cline{2-2}
{\color{black}}                                                                                               & {\color{black}\begin{tabular}[c]{@{}l@{}}\-- Spectral efficiency gains analysis under three types of rate selection/adaptation conditions, including \\ ~{} Fixed MCS, Variable MCS, Feedback Based Rate Adaptation {[}103{]}\end{tabular}}                                        \\ \cline{2-2}
\multirow{-6}{*}{{\color{black} \textbf{\begin{tabular}[c]{@{}c@{}}Performance \\ Analysis\end{tabular}}}}     & {\color{black}\begin{tabular}[c]{@{}l@{}}\-- Goodput performance analysis of IEEE 802.11be with HARQ in collision-free (AWGN-impaired) \\ ~{} and collision-dominated (interference-impaired) environments {[}104{]}\end{tabular}}                                                 \\ \hline
\end{tabular}
\end{table*}

\section{Enhanced Link Adaptation And Retransmission}
HARQ is another candidate feature under discussions for EHT, which combines retransmissions before decoding to improve performance (e.g., increasing reliability, reducing latency). So far, several issues regarding HARQ, such as HARQ granularity, HARQ process, and HARQ methods, have been raised and summarized in Table VIII. In this section, we elaborate on intertwined aspects of the PHY and MAC which need to be designed better for successful support of HARQ in EHT.

\subsection{HARQ Granularity}
The granularity represents an error-check granularity for retransmissions. In current IEEE 802.11 systems, fine ARQ granularity can only be supported at the MAC level, but not at the PHY level. Theoretically, as shown in Fig. 19, HARQ granularity can be supported at the A-MPDU level, MPDU level and CW level \cite{ref93}. For different levels of HARQ retransmissions, we will discuss what possible changes would be required at the PHY and MAC layer.

\subsubsection{HARQ at A-MPDU Level}
When the whole A-MPDU in Fig. 19(a) is to be retransmitted, the retransmitted A-MPDU usually has changes as compared to the original transmission because of: arbitrary numbers of delimiters among the MPDUs, each MPDU header's retry bit, different ciphertext, and different CRC bits \cite{ref93}. Due to these few different bits in the MAC payload resulting in different payload at PHY, combining of log-likelihood ratios(LLRs) at the PHY on the transmission will not be possible. Besides, at the PHY, there is no knowledge of MPDUs, and this payload is transformed into retransmitted CWs that are definitely different from CWs corresponding to the original transmission. Thus, changes at MAC may be needed to ensure that the same A-MPDU as from the original transmission is retransmitted.

\subsubsection{HARQ at MPDU Level}
In Fig. 19(b), it can be observed that the failed MPDU can span two partial CWs and one complete CW. The failed MPDU will have the different bits in the retransmission: MPDU header's retry bit, ciphertext and CRC bits \cite{ref93}. Thus, the CWs corresponding to the failed MPDU has the different coded bits and cannot be combined at LLR level. Besides, this failed MPDU will be mapped onto different CWs, and each CW will have different FEC padding. The misalignment in the failed and retransmitted MPDU's CWs will lead to the failed CW-combining at PHY.
\begin{figure*}[!htbp]
  \begin{center}
    \setlength{\abovecaptionskip}{-0.5cm}
    \scalebox{1}[1]{\includegraphics{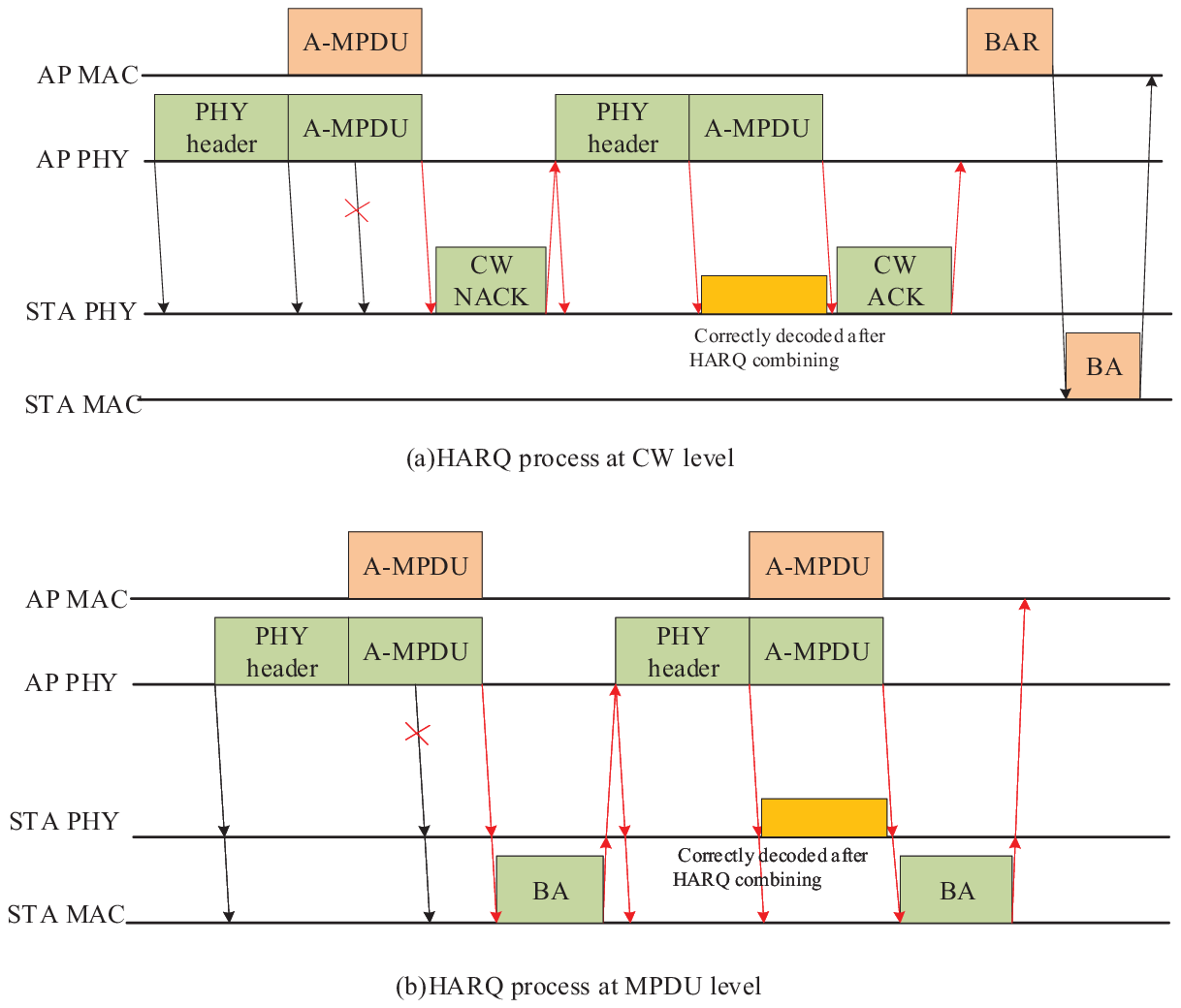}}
    \renewcommand{\figurename}{Fig.}
    {\color{black}\caption{An example of the HARQ process. (a) For the HARQ process at the CW level, it is determined whether or not the whole A-MPDU has been successfully transmitted at the PHY layer. Error checking of the retransmitted A-MPDU is performed on a per-CW(s) basis, and a new feedback mechanism and a new HARQ feedback frame (e.g., CW NACK/ACK) should be defined. (b) For the HARQ process at the MPDU level, when the whole A-MPDU is to be retransmitted, the MAC layer prepares the A-MPDU to be retransmitted. Error checking is performed on an MPDU basis, and the existing BA protocol can be used as the HARQ feedback.}}
  \end{center}
\end{figure*}
\subsubsection{HARQ at CW Level}
In current IEEE 802.11 standards, there is no support for PHY level retransmission as well. In Fig. 19(c), HARQ requires the PHY to recognize erroneous CWs so that it can combine the retransmitted CWs. Therefore, more signaling will be involved in the PHY to indicate the mapping between CWs and MPDUs and associate failed CWs to MPDUs. Additionally, the MAC layer needs to inform the PHY by checking CRC about the status of MPDUs. Each CW may need extra CRC information to prevent false detection \cite{ref94}, such as Forward Error Correction appending at the end of each CW.\par
The misalignment of CWs in the retransmitted A-MPDU or MPDU is a significant issue that poses great challenges to combining and decoding operations. To align the border of MPDU and CWs, the bitstream at the output of the decoder can be presented to the MAC for parsing, identifying delimiters and verifying the CRCs of any subframes \cite{ref93}. In this way, it allows the MAC layer to determine which MPDUs are received correctly. At the same time, the PHY can identify the locations of the CWs that are not decoded correctly. As shown in Fig. 19(d), MAC padding discussed in \cite{ref93}\cite{ref94}\cite{ref95} can also be adapted to align the border of MPDU and CW while maintaining the existing LDPC and BA designs. In this method, necessary mapping information between MPDU and CW shall provide the minimum information to the receiver for combining the required CWs. However, padding can incur additional padding overhead.

\subsection{HARQ Process}
An HARQ-capable STA attempting to decode a retransmitted PPDU does not ignore the previous unsuccessful PPDU but instead combines their bits by LLRs to improve the likelihood of correct decoding. HARQ is implemented on the data portion of the PPDU and needs knowledge of the parameters of the PPDU to identify a PPDU that should be combined. The parameters may be carried in the PHY header \cite{ref97}, a new SIG field in the PHY/MAC layer \cite{ref94,ref97,ref98,ref99}, or a new MAC frame \cite{ref97}, which are usually encoded separately from the data using different transmission parameters (e.g., MCS, coding rate) and have better decoding performance. According to different retransmission granularity, the PHY-level HARQ process based on CWs and the MAC-level HARQ process based on MPDUs are considered in \cite{ref95}, as shown in Fig. 20. The error check is performed on a per-codeword(s) basis by the parity check in the PHY-level process and is performed by the CRC check on an MPDU basis in the MAC-level process. To initiate the HARQ process, it is necessary that a HARQ feedback method to indicate the reception check status and which CWs/MPDUs are required to be retransmitted, such as using the existing BA frame as MAC-level HARQ feedback with minimal design effort and designing a new PHY-level feedback frame including sequence number for tracking each CW. In addition, the information from MAC (MPDU) and PHY (CWs) can now be combined into a single BA frame that acknowledges the reception of MPDUs and requests retransmission of CWs \cite{ref93}. Such dual feedback may be more efficient than creating a new HARQ feedback frame applied to the CW level.\par
Based on the HARQ feedback, as shown in Fig. 20(a) the STA initiates the HARQ operation at PHY for each transmission until it is either successful or reaches maximum HARQ retransmissions. The STA can use parity check to determine if it needs to store this CW or not, and then sends back the feedback about failed CWs to the AP. The AP needs to store the original CWs at the PHY level and retransmit only the CWs which are asked for in the HARQ feedback. The STA receives these retransmitted CWs and combines them with previous stored CWs, and forwards the correctly decoded bitstream to the MAC layer. The MAC layer performs the CRC check and terminates the retransmission by sending an MPDU level feedback to the AP. The HARQ session to be finished before a new transmission can be started on the medium. Since there can be retransmissions between PHYs, MAC layer may wait for a long time while their ARQ procedures cannot be supported for HARQ. Thus, it can be possible to define the fixed number of max retransmissions or redefine MAC timeout as mentioned in \cite{ref97}. In Fig. 20 (b), MPDU and CW can be aligned by using additional MAC padding so that the retransmission procedure can be relatively simple. However, it can cause additional overhead. Without additional padding, mapping information between MPDU and CW can cause additional overhead. In addition, it can be observed that MAC-level retransmission is more compatible with the existing ARQ than PHY-level retransmission.

\subsection{HARQ Methods}
In general, there are three methods for HARQ retransmission \cite{ref100}\cite{ref101}, CC which retries the same coded MPDU, punctured CC (PCC) which is modified upon CC, and IR which retries with additional parity. {\color{black} With CC, since the retransmission sends the same code bits of the CW corresponding to the MPDU of the initial transmission, it is easy to combine the retransmitted bits with the stored bits. CC requires BCC interleaver or LDPC tone mapper varying over different transmissions to achieve the frequency diversity gain, while PCC and IR have no such requirements. With PCC, all code bits of the LDPC are transmitted in the initial transmission, while the punctured code bits are retransmitted in the HARQ retransmissions. Similar to the PCC, IR supports puncturing of code bits of a CW in the HARQ retransmission. In the initial transmissions, all the information bits and a subset of the parity bits will be transmitted, while the bits containing all the information that are punctured (not transmitted) in the initial transmission are transmitted in the retransmissions. Unlink CC, IR has some other issues due to the coding method. For BCC, using different puncturing patterns in each retransmission can be considered to make new parity bits, which is only applied to RU size less than or equal to 242 tones. Hence the gain cannot be realized for larger RU sizes. For LDPC encoding scheme, IR can support current LDPC code rates (1/2, 2/3, 3/4, 5/6) and CW length (1944 bits) employed in current IEEE 802.11 standards \cite{ref100}\cite{ref101}.} Different HARQ methods have different implementation complexities and gains. How to choose the exact HARQ methods in different environments is challenging. The gain from HARQ is dependent on the link adaptation method \cite{ref102} which tries to change the transmission rate based on the real-time channel status to improve the transmission performance before the initial transmission. There are three types of rate selection/adaptation techniques: fixed MCS in the initial transmission and retransmission, variable MCS based on long-term SNR, and optimal rate adaptation based on HARQ feedback, which can include information like the short-term SNR. HARQ (PCC and IR) spectral efficiency gains under three types of rate selection/adaptation conditions were investigated in \cite{ref103}, and the simulation results showed that the largest gains are for the case of feedback-based rate adaption and below around 25 dB SNR and losses due to overhead become significant for shorter HARQ PPDU durations. Thus, it will be essential to limit overhead loss to realize HARQ gains in real systems. In \cite{ref103}, the goodput performance of EHT with HARQ in collision-free (additive white Gaussian noise-impaired) and collision-dominated (interference-impaired) environments was evaluated. Simulation results also demonstrated performance gains but showed that there is a need for different strategies in collision-dominated environments. Consider more various conditions (e.g., hardware complexity), further study on implementing HARQ in typical HARQ scenarios is still needed, e.g., multiple links, multiple users or multiple APs scenarios.
\subsection{Summary of the HARQ}
This section presents several issues and challenges regarding HARQ granularity, HARQ process and methods. Theoretically, the granularity and process of HARQ can be supported at the MAC layer level and the PHY level. The basic HARQ granularity in EHT is not determined yet. As aforementioned, for the A-MPDU-level/MPDU-level retransmissions, HARQ combining is ineffective when the retransmitted bitstream differs from the initial transmitted bitstream. Concerning the CW-level retransmission, the HARQ combining that only requires minor changes and designs based on the existing specifications, which is therefore relatively simple. By contrast, using the CW-level retransmission is considered more feasible since the drawback of using the A-MPDU-level/MPDU-level retransmission without one-to-one mapping between the A-MPDU/MPDU and CWs can be relatively hard to overcome. However, EHT requires a more detailed and in-depth investigation and analysis for the standardization of HARQ, such as CWs processing and memory requirements for erroneous CWs. As in Section VI, the PHY-level retransmission process and the MAC-level retransmission process are investigated. By contrast, in terms of signaling and potential padding overhead, the PHY-level retransmission is preferred while we need to design new feedback and consider how to cooperate with the existing ARQ protocol. In the case of the MAC-level retransmission, signaling overhead to map MPDUs with CWs and additional MAC padding overhead will be significant. Thus, we think that using PHY-level retransmission is appropriate since the drawback of using PHY-level retransmission with lower PHY-MAC interaction overhead can be relatively easily resolved.

\section{Future Development and Research Opportunities}
In this section, some open technical issues that need to be investigated and several promising research directions are discussed. These topics are expected to promote the development of wireless communication networks by addressing several technical challenges including 6GHz co-existence problems, integrating low-frequency and high-frequency bands, guaranteed QoS provisioning based on machine learning, power management and hybrid beamforming.
\subsection{Coexistence in the 6 GHz band}
One of the main objectives of EHT is to make full use of up to 1.2 GHz spectrum resources in the 6 GHz band. However, to effectively utilize these frequency resources, EHT has to co-exist with other different technologies operating in the same band, such as IEEE 802.11ax and 5G on the unlicensed band. Coexistence among wireless networks is challenging, especially when these networks are heterogeneous. The spectrum access rules differ across networks, which may hinder the fair sharing of the spectrum resources. We should distinguish at least two coexistence scenarios \cite{ref105}: one is that networks implement coexistence mechanisms on its own without any consultation from its neighbours, and the other is that networks directly or indirectly coordinate to ease their coexistence. In a coordinated coexistence setting, coordination may require a common management/control plane between heterogeneous technologies, e.g., control of cloud \cite{ref63}, SDN \cite{ref106}, or a direct communication tunnel. In an uncoordinated coexistence setting, uncoordinated schemes might require more sophisticated techniques to implement neighbour-aware coexistence schemes, and networks try to ensure coexistence with other networks mostly based on their local observations. Machine learning has been envisaged to be critical for reaching coexistence goals by providing the necessary intelligence and adaptation mechanisms, e.g., using machine learning for coexistence by discovering network dynamics \cite{ref106} or predicting spectrum usage of adjacent networks \cite{ref107}. New machine learning algorithms will fast learning speed while leveraging distributed computing resources over edge and cloud could be extremely useful in achieving situation-aware coexistence.

\subsection{Integrating low-frequency and high-frequency bands}
Densely deployed sub-6GHz WLANs alone may not provide the seamless connectivity required by mobile services and the rapid increase in mobile data traffic in future wireless networks. As a result, one of the main advancements in the network design for WLAN relies on the integration of multiple different bands (e.g., microwave and mmWave). An integrated system that can leverage multiple frequencies across the microwave/mmWave/THz spectrum is needed to provide seamless and intelligent connectivity at both wide and local area levels. The existing IEEE 802.11 standards can already provide a negotiation pipe among different frequency bands through FST and On-channel tunneling for multi-band operations (e.g., fast session transfer) \cite{ref7}. Besides, there are already many potential solutions for mobility management and network data migration management to integrate the microwave/mmWave spectrum \cite{ref109,ref110,ref111,ref112,ref113}. However, exploiting integrated low and higher frequency bands will bring forth several new open problems from hardware to system design. For example, supporting high mobility at these multiple spectra \cite{ref114}\cite{ref115}, developing new multiple access and networking paradigms, and new transceiver architectures are needed along with new microwave/mmWave/THz frequencies propagation models. Another important research direction is to study collaborative operation across these multiple spectra.

\subsection{Guaranteed QoS provisioning based on machine learning}
How to intelligently and appropriately recognize the diverse QoS requirements and efficiently allocate wireless resources for users with different QoS requirements is a hot topic in the standardization process. In the existing QoS-supported WLAN, EDCA adjusts back-off parameters to implement priority-based channel access at the MAC layer \cite{ref7}, which can provide a certain degree of QoS guarantee for different types of traffics with different QoS requirements. The appearance of emerging and diverse network services (e.g., gaming, VR/AR, and 4k/8k video) has created new challenges in the diverse QoS requirements of different users, such as throughput, delay, jitter, and loss rate. This often requires the network to have to react quickly to user's experience, allocate wireless resources for users with different QoS requirements, and offer a better QoS. However, current EDCA is good in the view of statistics, but may be unable to help improve the worst-case latency and jitter due to such limitations, e.g., discriminating different types of traffics only according to QoS fields and not directly reflecting latency requirements of latency-sensitive applications. In addition to using the shortened Trigger frames for soliciting Trigger-based PPDU transmission carrying the real-time traffic \cite{ref116}, a new design of queue \cite{ref117} may improve the worst-case latency by (i) linking access delay directly to the latency required by the latency-critical traffic, (ii) categorizing traffic by taking into consideration more parameters, and (iii) being grant with higher priority in channel access, etc. Besides, machine-learning methods with situation-awareness \cite{ref118} may be efficient ways to satisfy users' QoS requirements by a set of observations reflecting the network state and the user's perception. For example, a multi-link capable device can apply machine-learning algorithms to predict the future link status (busy/idle) according to a set of historical link information (e.g., link status, link traffic or link utilization), and then adaptively and quickly switch the link subjected to strong interference to another appropriate link so as to guarantee the communication quality. In addition, machine-learning algorithms are currently considered as promising tools for various purposes in many fields ranging from PHY/MAC protocols design to the development of theoretical foundations, such as parameters optimization (e.g., adjusting contention window or priority for EDCA), protocol version selection, multi-channel/multi-link aggregation, channel modeling, fast time-varying channel estimation, modulation recognition, RU allocation, multi-antenna selection with 16 spatial streams for SU-MIMO/MU-MIMO transmission, multi-AP selection and multi-AP network deployment for multi-AP coordination, etc. For example, a multi-AP network can predict the future users' QoS requirements and environment conditions in terms of properties such as the multi-AP transmission parameters/settings (e.g., MSC/bandwidth), and then optimize the network resource management and improve the overall multi-AP network performance.
\subsection{Power management}
Mobile devices are battery-powered and have limited battery life. In addition to improving the design of the battery itself, it is critical to enhance energy-saving mechanisms. With the new requirements and characteristics of EHT, such as multi-link operation, multi-AP collaboration, and HARQ, the power consumption level of EHT devices could be significantly increased. For example, in light traffic load conditions, a multilink-capable mobile device may be in listening mode for quite long time, which may constitute a significant ratio of a multi-link device's power consumption. As a result, new and more effective energy-saving mechanisms should be carefully designed to cope with the increased power consumption. For example, flexibly enabling/disabling links based on the actual network conditions or effective prediction methods may be an ideal approach to save a lot of power compared to the modes where all links are involved in. Also, the combination of artificial intelligence algorithms and power management to achieve intelligent high efficiency and energy saving \cite{ref119}\cite{ref120} is also a direction worthy of research. For example, in a multi-AP network, the APs participating in the multi-AP transmissions can intelligently increase or decrease the transmit power based on the accurately predicted movement trajectory of users, user's QoS requirements and channel conditions. Nevertheless, the timeliness and accuracy of prediction algorithms for power saving still need to be deeply investigated.

\subsection{Hybrid beamforming}
Unlike digital beamforming currently used in sub-6GHz WLANs, where every spatial stream necessitates an expensive RF chain \cite{ref7}, hybrid beamforming often requires that the number of RF chains may be much smaller than the actual number of antennas. Hybrid beamforming can help balance flexibility and cost trade-offs while still fielding a system that meets the required performance parameters. To achieve hybrid beamforming in future wireless communications systems, the main issue to be considered are the system models of transceivers' structures and the matrices with the possible antenna configuration scenarios \cite{ref121}. And, a system-level model of hybrid beamforming and modeling algorithms should be explored and assessed over a collection of parameters (e.g., RF, antenna, and signal processing components), steering, and channel combinations.

\section{Conclusions}
In recent years, some new applications with ultra-high throughput and ultra-low latency requirements have prompted the IEEE 802.11 standard to evolve further to accommodate these new services features. For example, VR, social networking, the internet of things, and ultra-high-speed content delivery place challenging WLAN requirements. As a result, the IEEE 802.11 standard will continue to conduct evolution on key techniques and revision of the new standard known as EHT or in the new wording called Wi-Fi 7. Undoubtedly, designing high-performance PHY and MAC protocol for EHT is a challenging task, but it is also an exciting research area. In this survey, we focus on open issues and crucial proposals proposed by us and others in the standardization process of EHT and give some free topics. Concretely, this article elaborates on the most attractive techniques that may be written into EHT WLAN standard, including multi-RU support, 4096-QAM, multi-link aggregation and operations, MIMO enhancements, multi-AP coordination techniques, and HARQ. Clearly, such tremendous changes from the existing WLAN protocol could make the EHT be a landmark standard protocol in the evolution of the IEEE 802.11 family. Further, some free research perspectives, related to the 6 GHz co-existence, integrating low-frequency and high-frequency bands, guaranteed QoS provisioning based on machine learning, hybrid beamforming, and power management, have been discussed briefly. Besides, the peak rate of the PHY layer could be enhanced by using other more efficient encoding technologies and new multiple access technologies (e.g., non-orthogonal multiple access technologies). Furthermore, as an emerging and attractive field in recent years, Wi-Fi sensing for low-cost and low-complexity hand gesture recognition, high-precision motion detection, health monitoring, ranging and positioning and so on, which has attracted wide attention and will be incorporated into the IEEE 802.11 protocol. As you can see and hear, the EHT standardization process has just started, and most of the researches is underway and open. Consequently, we hope that this article could draw researchers' attention to EHT and thus promote the development of future WLANs.

\section*{Appendix A}
\begin{center}
Summary of Main Acronyms.
\end{center}
\begin{footnotesize}
\arrayrulecolor{black}
\begin{center}
\tablefirsthead{\hline\multicolumn{1}{|c|}{\textbf{Acronyms}}&\multicolumn{1}{|c|}{\textbf{Definition}}\\}
\tablelasttail{\hline}
\tablehead{\hline\multicolumn{2}{|c|}
{\small\sl Continued from previous page}\\\hline
\multicolumn{1}{|c|}{\textbf{Acronyms}}&\multicolumn{1}{|c|}{\textbf{Definition}}\\}
\tabletail{\hline\multicolumn{2}{|c|}{\small\sl Continued on next page}\\\hline}
\begin{supertabular}{!{\vrule width0.6pt}c!{\vrule width0.6pt}p{6.2cm}<{\centering}!{\vrule width0.6pt}}
\Xhline{0.6pt}
AC & Access Controller \\
\Xhline{0.6pt}
ACK & Acknowledgment \\
\Xhline{0.6pt}
A-MPDU & Aggregate MPDU \\
\Xhline{0.6pt}
AP & Access Point \\
\Xhline{0.6pt}
ARQ & Automatic Repeat Request \\
\Xhline{0.6pt}
CBF & Coordinated Beamforming  \\
\Xhline{0.6pt}
CC & Chase Combining\\ \hline
CCA & Clear Channel Assessment \\
\Xhline{0.6pt}
CFO & Carrier Frequency Offsets \\
\Xhline{0.6pt}
C-OFDMA & Coordinated Orthogonal Frequency-Division Multiple Access \\
\Xhline{0.6pt}
CRC & Cyclic Redundancy Check  \\
\Xhline{0.6pt}
CSI & Channel State Information  \\
\Xhline{0.6pt}
CSMA & Carrier Sense Multiple Access \\
\Xhline{0.6pt}
CSMA/CA & Carrier Sense Multiple Access/Collision Avoidance \\
\Xhline{0.6pt}
CSR & Coordinated Spatial Reuse \\
\Xhline{0.6pt}
CW & Codeword \\
\Xhline{0.6pt}
DL & Downlink  \\
\Xhline{0.6pt}
EDCA & Enhanced Distributed Channel Access \\
\Xhline{0.6pt}
EHT & Extremely High Throughput \\
\Xhline{0.6pt}
EVM & Error Vector Magnitude  \\
\Xhline{0.6pt}
FST & Fast Session Transfer\\
\Xhline{0.6pt}
HARQ & Hybrid Automatic Repeat Request\\
\Xhline{0.6pt}
IEEE & Institute of Electrical and Electronic Engineers  \\
\hline
IR & Incremental Redundancy \\
\Xhline{0.6pt}
JXT & Joint Transmission \\
\Xhline{0.6pt}
LDPC & Low Density Parity Check \\
\Xhline{0.6pt}
LLR & Log-likelihood Ratio \\
\Xhline{0.6pt}
MAC & Medium Access Control\\
\Xhline{0.6pt}
M-AP & Master AP  \\
\Xhline{0.6pt}
MCS & Modulation Coding Scheme\\
\Xhline{0.6pt}
MIMO & Multiple Input Multiple Output \\
\Xhline{0.6pt}
MLME & Mac Sublayer Management Entity \\
\Xhline{0.6pt}
MPDU & Mac Protocol Data Units \\
\Xhline{0.6pt}
MSDU & Mac Service Data Unit\\
\Xhline{0.6pt}
MU-MIMO & Multi-User MIMO \\
\Xhline{0.6pt}
NDP & Null Data Packet \\
\Xhline{0.6pt}
NDPA & Null Data Packet Announcement\\
\Xhline{0.6pt}
nonCTX & Non-Coordinated Transmission \\
\Xhline{0.6pt}
OFDMA & Orthogonal Frequency Division Multiple Access\\
\Xhline{0.6pt}
PCC & Punctured Chase Combining \\
\Xhline{0.6pt}
PHY & Physical Layer \\
\Xhline{0.6pt}
PN & Packet Number \\
\Xhline{0.6pt}
PPDU & Physical Protocol Data Unit \\
\Xhline{0.6pt}
PU & Processing Unit \\
\Xhline{0.6pt}
QAM & Quadrature Amplitude Modulation \\
\Xhline{0.6pt}
QoS & Quality of Service \\
\Xhline{0.6pt}
RF & Radio Frequency \\
\Xhline{0.6pt}
RSNA & Robust Security Network Association \\
\Xhline{0.6pt}
RU & Resource Unit \\
\Xhline{0.6pt}
SAP & Service Access Points \\
\Xhline{0.6pt}
SDN & software-defined network \\
\Xhline{0.6pt}
S-AP & Slave AP \\
\Xhline{0.6pt}
SINR & Signal to Interference plus Noise Ratio \\
\Xhline{0.6pt}
SME & Station Management Entity \\
\Xhline{0.6pt}
SNR & Signal-Noise Ratio \\
\Xhline{0.6pt}
STA & Station \\
\Xhline{0.6pt}
SU & Single-User \\
\Xhline{0.6pt}
T-PCH & Temporary Primary Channel \\
\Xhline{0.6pt}
TXOP & Transmission Opportunity \\
\Xhline{0.6pt}
UL & Uplink  \\
\Xhline{0.6pt}
Wi-Fi & Wireless Fidelity \\
\Xhline{0.6pt}
WLAN & Wireless Local Area Network \\
\Xhline{0.6pt}
\end{supertabular}
\end{center}
\end{footnotesize}

\ifCLASSOPTIONcaptionsoff
  \newpage
\fi

\end{document}